\documentclass{aa}

\usepackage{graphicx}
\usepackage{txfonts}
\usepackage{amssymb}
\usepackage[T1]{fontenc}
\usepackage{textcomp}
\usepackage{longtable}
\usepackage{booktabs}
\usepackage{xcolor}
\usepackage{lscape}
\usepackage[switch]{lineno}

\DeclareUnicodeCharacter{2212}{-}

\usepackage{natbib}
\usepackage{hyperref}
\begin{document} 


\title{SN\,2020bqj: a Type Ibn supernova with a long lasting peak plateau}
\subtitle{}

\author{E. C. Kool\inst{1}
\and{E. Karamehmetoglu}\inst{2}
\and{J. Sollerman}\inst{1}
\and{S. Schulze}\inst{3}
\and{R. Lunnan}\inst{1}
\and{T. M. Reynolds}\inst{4}
\and{C. Barbarino}\inst{1}
\and{E. C. Bellm}\inst{5}
\and{K. De}\inst{6}
\and{D. A. Duev}\inst{7}
\and{C. Fremling}\inst{7}
\and{V.~Z. Golkhou}\inst{5,8}
\and{M. L. Graham}\inst{5}
\and{D. A. Green}\inst{9}
\and{A. Horesh}\inst{10}
\and{S. Kaye}\inst{11}
\and{Y.-L. Kim}\inst{12}
\and{R. R. Laher}\inst{13}
\and{F. J. Masci}\inst{13}
\and{J. Nordin}\inst{14}
\and{D. A. Perley}\inst{15}
\and{E.~S. Phinney}\inst{16}
\and{M. Porter}\inst{11}
\and{D. Reiley}\inst{11}
\and{H. Rodriguez}\inst{11}
\and{J. van Roestel}\inst{7}
\and{B. Rusholme}\inst{13}
\and{Y. Sharma}\inst{7}
\and{I. Sfaradi}\inst{10}
\and{M. T. Soumagnac}\inst{17,3}
\and{K. Taggart}\inst{15}
\and{L. Tartaglia}\inst{1}
\and{D. R. A. Williams}\inst{18}
\and{L. Yan}\inst{6}
}

\institute{
The Oskar Klein Centre, Department of Astronomy, Stockholm University, AlbaNova, SE-10691, Stockholm, Sweden
\email{erik.kool@astro.su.se}
\and{Department of Physics and Astronomy, Aarhus University, Ny Munkegade 120, DK-8000 Aarhus C, Denmark} 
\and{Department of Particle Physics and Astrophysics, Weizmann Institute of Science, Rehovot 7610001, Israel} 
\and{Tuorla observatory, Department of Physics and Astronomy, University of Turku, FI-20014 Turku, Finland} 
\and{DiRAC Institute, Department of Astronomy, University of Washington, 3910 15th Avenue NE, Seattle, WA 98195, USA} 
\and{Cahill Center for Astrophysics, California Institute of Technology, 1200 E. California Blvd. Pasadena, CA 91125, USA}
\and{Division of Physics, Mathematics, and Astronomy, California Institute of Technology, Pasadena, CA 91125, USA}
\and{The eScience Institute, University of Washington, Seattle, WA 98195, USA}
\and{Astrophysics Group, Cavendish Laboratory, 19 J. J. Thomson Avenue, Cambridge CB3 0HE, UK}
\and{Racah Institute of Physics, The Hebrew University of Jerusalem, Jerusalem, 91904, Israel}
\and{Caltech Optical Observatories, California Institute of Technology, Pasadena, CA  91125, USA} 
\and{Universit\'e de Lyon, Universit\'e Claude Bernard Lyon 1, CNRS/IN2P3, IP2I Lyon, F-69622, Villeurbanne, France} 
\and{IPAC, California Institute of Technology, 1200 E. California Blvd, Pasadena, CA 91125, USA}
\and{Institute of Physics, Humboldt-Universit\"{a}t zu Berlin, Newtonstr. 15, 12489 Berlin, Germany} 
\and{Astrophysics Research Institute, Liverpool John Moores University, 146 Brownlow Hill, Liverpool L3 5RF, UK} 
\and{Theoretical Astrophysics, 350-17 California Institute of Technology, Pasadena CA 91125, USA} 
\and{Lawrence Berkeley National Laboratory, 1 Cyclotron Road, Berkeley, CA 94720, USA} 
\and{Jodrell Bank Centre for Astrophysics, School of Physics and Astronomy, The University of Manchester, Manchester, M13 9PL, UK} 
}

\date{}
 
\abstract
   {Type Ibn supernovae (SNe Ibn) are a rare class of stripped envelope supernovae interacting with a helium-rich circumstellar medium (CSM). The majority of the SNe Ibn reported in the literature display a surprising homogeneity in their fast-evolving lightcurves and are typically found in actively starforming spiral galaxies.}
   {We present the discovery and the study of SN\,2020bqj (ZTF20aalrqbu), a SN Ibn with a long-duration peak plateau lasting 40 days and hosted by a faint low-mass galaxy. We aim to explain its peculiar properties using an extensive photometric and spectroscopic data set.}
   {We compare the photometric and spectral evolution of SN\,2020bqj with regular SNe Ibn from the literature, as well as with other outliers in the SN Ibn subclass. We fit the bolometric and multi-band lightcurves with powering mechanism models such as radioactive decay and CSM interaction. We also model the host galaxy of SN\,2020bqj.}
   {The risetime, peak magnitude and spectral features of SN\,2020bqj are consistent with those of most SNe Ibn, but the SN is a clear outlier in the subclass based on its bright, long-lasting peak plateau and the low mass of its faint host galaxy. We show through modeling that the lightcurve of SN\,2020bqj can be powered predominantly by shock heating from the interaction of the SN ejecta and a dense CSM, combined with radioactive decay. The peculiar Type Ibn SN\,2011hw is a close analog to SN\,2020bqj in terms of lightcurve and spectral evolution, suggesting a similar progenitor and CSM scenario. In this scenario a very massive progenitor star in the transitional phase between a luminous blue variable and a compact Wolf-Rayet star undergoes core-collapse, embedded in a dense helium-rich CSM with an elevated opacity compared to normal SNe Ibn, due to the presence of residual hydrogen. This scenario is consistent with the observed properties of SN\,2020bqj and the modeling results.}
   {SN\,2020bqj is a compelling example of a transitional SN Ibn/IIn based on not only its spectral features, but also its lightcurve, host galaxy properties and the inferred progenitor properties. The strong similarity with SN\,2011hw suggests this subclass may be the result of a progenitor in a stellar evolution phase that is distinct from those of progenitors of regular SNe Ibn.}


\maketitle

\section{Introduction} \label{sec:intro}
Supernovae of Type Ibn (SNe Ibn) are a rare class of core-collapse SNe, with less than 40 objects reported in the literature \citep{hosseinzadeh2019}. SNe Ibn are classified based on spectra that are dominated by relatively narrow ($\sim$few$\times 10^{3}$ km s$^{-1}$, hence the `n' suffix) helium (He) emission lines while showing little to no hydrogen (H). The intermediate-width emission lines are interpreted as a sign of shock interaction between the fast-moving SN ejecta and a slow-moving He-rich, but H-depleted, circumstellar medium (CSM) \citep[for a review see][]{smith2017}. The He-rich CSM around SNe Ibn is assumed to have originated from their progenitor stars through mass-loss, as is known to be the case for their Type IIn SN counterparts (SNe with spectra showing narrow H emission lines) based on observations of pre-SN outbursts \citep{fraser2013,ofek2013,ofek2014,strotjohann2020}. As such, SNe Ibn are expected to be stripped-envelope SNe (SE SNe) embedded in a He-rich environment \citep{pastorello2008I,foley2007,chugai2009}. 

Based on the He-rich CSM around SNe Ibn, with unperturbed CSM velocities of $\sim$500-1500 km s$^{-1}$ \citep{pastorello2016}, the progenitors of SNe Ibn are commonly assumed to be massive evolved Wolf-Rayet (WR) stars, which have atmospheres that are nearly H-free and exhibit mass-loss through strong and fast winds 
\citep{prinja1990,crowther2007}. Furthermore, virtually all SNe Ibn are observed to occur in star-forming galaxies \citep{pastorello2015VI}, supporting the association of SNe Ibn to short-lived massive progenitors such as WR stars. 

In so-called interacting SNe such as SNe Ibn, the shock interaction between ejecta and CSM converts kinetic energy to visible light, which means CSM interaction contributes to (or even dominates) the observed energy output of an interacting SN. In Type IIn SNe this results in a large diversity in lightcurves \citep[e.g.,][]{taddia2013,nyholm2020}, due to variation in the properties (composition, density, mass, geometry) of the CSM \citep{smith2017,Soumagnac2019}. In contrast, the lightcurves of SNe Ibn are surprisingly homogeneous. They are typically fast and evolve monotonically with a risetime of $\lesssim$ 15 days to a peak absolute magnitude of $M_r$ $\approx -19$, followed by a decline at a rate of $\sim0.05 - 0.15$ mag day$^{-1}$ during the first month after peak \citep{hosseinzadeh2017}. Normal SE SNe (Type Ib/c) have lightcurves that are generally consistent with powering by the decay of radioactive material synthesized in the explosion and present in the ejecta ($^{56}$Ni $\rightarrow$ $^{56}$Co $\rightarrow$ $^{56}$Fe, e.g., \citealt{prentice2016}). In the case of SNe Ibn, however, the high peak luminosity and the rapid lightcurve evolution typically exclude $^{56}$Ni decay as the main powering mechanism for their lightcurves, as the decline rate is usually too steep to reconcile with the amount of $^{56}$Ni required to explain the peak luminosity \citep[e.g.,][]{moriya2016}. Instead, models combining radioactive decay and interaction with a (He-rich) CSM shell have been used to reproduce the observed fast-evolving SN Ibn lightcurves \citep{karamehmetoglu2019,wang2019,clark2020,gangopadhyay2020}. These model fits result in $^{56}$Ni and CSM masses that are consistent with WR progenitors, although they are affected by substantial uncertainties and degeneracy given the large number of free model parameters.

Despite the apparent match of the massive WR progenitor scenario with observed SN Ibn properties, there are still many open questions regarding the conditions and origin of the CSM, the homogeneity of SN Ibn progenitors, and the (dominant) powering mechanisms of SNe Ibn. Unlike Type IIn SNe where some have been linked to H-rich luminous blue variable (LBV) stars \citep[e.g.,][]{galyam2007,trundle2008,galyam2009,smith2010}, there have been no direct progenitor detections of SNe Ibn. The only direct evidence of a massive SN Ibn progenitor has been a luminous outburst of the prototypical Type Ibn SN\,2006jc, observed two years before explosion \citep{nakano2006}. In contrast with LBV stars, such eruptions are not common for H-deficient WR stars, so the precursor of SN\,2006jc was deemed to have been a massive star with residual LBV-like properties \citep{foley2007,pastorello2007}, which would also explain the presence of faint H emission in the spectra. The outburst is assumed to be the source of the He-rich CSM shell, rather than a steady WR wind, based on the temporal coincidence of the outburst and the SN, and the high mass-loss rate inferred from the distance and mass of the CSM \citep{anupama2009,sun2020}.

An alternative scenario for the outburst preceding SN\,2006jc involves an LBV companion in a binary system with the progenitor of SN\,2006jc \citep{pastorello2007,pastorello2008I}. In fact, there is evidence of a surviving binary companion at the position of SN\,2006jc \citep{maund2016}, but the composition of the CSM and the properties of the surviving companion do not appear consistent with an LBV star \citep{sun2020}. However, the detection of a potential binary companion at the position of SN\,2006jc does suggest an alternative progenitor channel for SNe Ibn. The binary channel allows the progenitor to be of lower mass, stripped from its H envelope through binary interaction, rather than through mass-loss by a massive progenitor. Such an older low-mass progenitor could also explain PS1-12sk, the only SN Ibn known so far that was located at a site with extremely limited local star formation (SF) \citep{hosseinzadeh2019}, in an elliptical cluster galaxy \citep{sanders2013}.

Finally, there are a few examples of SNe Ibn with lightcurves that do not fit well the SN Ibn lightcurve template from \citet{hosseinzadeh2017}, either by showing multiple peaks or a slower evolution. In the case of SN\,2011hw, a scenario of a progenitor transitioning between the LBV and WR phases has been suggested to explain its spectral features (such as Balmer emission), and its unusual long-lived lightcurve \citep{smith2012,pastorello2015IV}. SN\,2011hw represents a small subset of SNe Ibn with H in their spectra and atypical lightcurves that are classified as transitional Type Ibn/IIn SNe \citep{pastorello2015IV}. However, the slowly evolving SN Ibn OGLE-2014-SN-131 did not show strong Balmer lines  \citep{karamehmetoglu2017}, and there are numerous examples of `normal' fast-evolving SNe Ibn with (prominent) H in their spectra \citep[e.g.,][]{pastorello2015VI,karamehmetoglu2019}.

In this paper we report on the discovery and the extensive follow-up program of SN\,2020bqj, an outlier in the SN Ibn subclass in almost all aspects. SN\,2020bqj reached a peak magnitude of $M_r\sim-19.3$ in less than six days, which is consistent with the subclass, but instead of declining afterwards the SN stayed roughly constant in magnitude for $\sim$40 days, followed by a slow linear decline spanning 90+ days. Additionally, in contrast to normal SNe Ibn, SN\,2020bqj was not located in a star-forming spiral galaxy, but instead the host appears to be a faint low-mass galaxy. Finally, the spectrum of SN\,2020bqj shows, in addition to intermediate-width He in emission, prominent C\,{\sc ii} emission lines during the first half of its lightcurve, while at later epochs H$\alpha$, O\,{\sc i}, Mg\,{\sc ii} and Ca\,{\sc ii} gain prominence.

This paper is organized as follows: In Sect.~\ref{sec:data} the observations of SN\,2020bqj are described. Sections~\ref{sec:photom_analysis}, \ref{sec:spectra_analysis} and \ref{sec:host_analysis} include analyses of the lightcurve, the spectrum and the host galaxy properties, respectively, where we compare these properties to those of SNe Ibn from the literature. In Sect.~\ref{sec:powering} we investigate the powering mechanisms of SN\,2020bqj by fitting models to the lightcurve. In Sect.~\ref{sec:discussion} we discuss the implications of our analysis of SN\,2020bqj on its CSM composition and SN progenitor properties. Finally, in Sect.~\ref{sec:summary} we summarize our findings. Throughout this paper we adopt a flat cosmology with $H_0$ = 70 km s$^{-1}$ Mpc$^{-1}$ and $\Omega_M$ = 0.3. 

\section{Observations}  \label{sec:data}
\subsection{Discovery}
SN\,2020bqj (= ZTF20aalrqbu), located at R.A. = $15^{\rm h}$ $33^{\rm m}$ $40\fs48$ and Dec. = $+34^\circ$ $28'$ $44\farcs3$ (J2000.0), was first detected by the Zwicky Transient Facility \citep[ZTF;][]{bellm2019,graham2019} with the ZTF camera \citep{dekany2020} mounted on the Palomar 48 inch (P48) telescope on 2020 February 2 (MJD 58881.52) with host subtracted AB magnitudes of $g = 18.05\pm0.06$ and $r = 18.41\pm0.06$ (all magnitudes in this work are given in the AB system). The transient was discovered 
in the ZTF alert stream \citep{patterson2019} after passing a filter that searches for fast rising transients, and subsequent monitoring was coordinated through the GROWTH Marshal \citep{kasliwal2019}. The transient was reported to the Transient Name Server (TNS) on the same day \citep{discovery_report}, and based on a spectrum obtained with the SPectrograph for the Rapid Acquisition of Transients \citep[SPRAT;][]{Piascik2014} on the Liverpool Telescope \citep[LT;][]{steele2004}, it was classified as a Type Ibn SN on 2020 February 21 \citep{perley2020}. Previous to discovery the field was last observed by ZTF in $g$- and $r$-band 3.0 days before, where the SN was not detected with global upper limits of $g$ > 20.2 mag and $r$ > 20.0 mag.

\begin{figure}
\centering
    \includegraphics[width=\hsize]{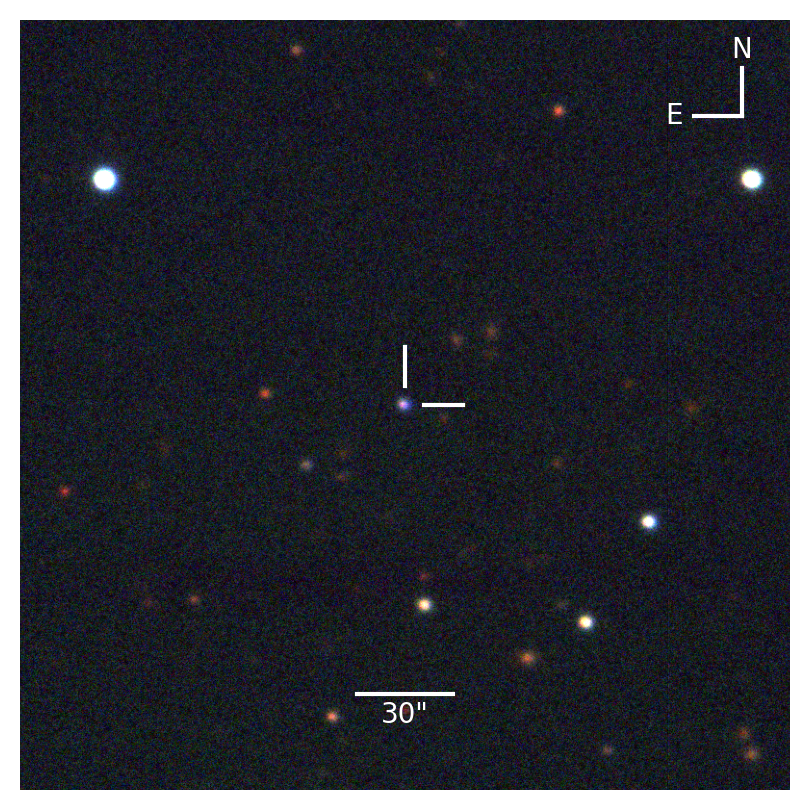}
    \caption{A $griz$-color composite image of SN\,2020bqj and its environment, as observed with LT/IO:O on 2020 March 6, +32 days after estimated explosion epoch. The SN outshines its faint host.}
    \label{fig:lt_griz}
\end{figure}

SN\,2020bqj appears hostless in our optical images, see Fig.~\ref{fig:lt_griz}, but a faint source ($r\sim23.5$) is reported in the eighth data release of the Dark Energy Spectroscopic Instrument (DESI) Legacy Survey \citep{dey2019}, at the same location as the SN and classified as a round exponential galaxy. No spectroscopic redshift has been reported for this source. The redshift towards the SN was determined based on seven intermediate-width (full-width half-maximum (FWHM) velocity of $\sim2000$ km s$^{-1}$) He lines and H$\alpha$ in the SN spectra, weighted by the respective uncertainty in each line center measurement, to be $z = 0.066 \pm 0.001$. Under the assumed cosmology this redshift corresponds to a luminosity distance of $d_L$ = 297 Mpc.

Milky Way (MW) extinction towards the SN is estimated to be $E(B-V)$ = 0.0194 mag \citep{schlafly2011}. All photometry are corrected for MW extinction adopting the Cardelli extinction law \citep{cardelli1989} with $R_V$ = 3.1. We assume host galaxy extinction towards the SN is negligible as the spectra do not show evidence of Na~\textsc{I D} absorption lines.

\subsection{Photometry}
\subsubsection{Optical}
Photometry was obtained with the ZTF camera mounted on the P48 telescope from nominal ZTF survey observations: the public Northern Sky Survey with a 3-day cadence in $g$- and $r$-band \citep{bellm2019b} and the ZTF Uniform Depth Survey (ZUDS, Goldstein et al., \textit{in prep}) at a nightly cadence in $g$-, $r$- and $i$-band from MJD 58954 onwards. Additional photometry was obtained from Palomar with the Spectral Energy Distribution Machine \citep[SEDM;][]{blagorodnova2018} mounted on the Palomar 60 inch telescope \citep[P60;][]{cenko2006}. The P48 data were reduced using the ZTF pipeline \citep{masci2019} and image subtraction based on the \citet{Zackay2016} algorithm, which produces template subtracted PSF photometry in the Sloan Digitial Sky Survey (SDSS) photometric system  calibrated against field stars selected from the Pan-STARRS1 survey \citep{chambers2016}. The P60 photometry was produced using the pipeline described in \citet{fremling2016}. Additional optical photometry in $ugriz$ was obtained with the Infrared-Optical imager (IO:O) on the LT. The observed optical photometry is listed in Table \ref{tab:photom} and the lightcurve, corrected for MW extinction, is shown in Fig.~\ref{fig:lightcurve}. The phase presented in the lightcurve figure and this work, unless otherwise stated, is rest-frame days since the estimated explosion epoch, which we adopt to be halfway between the last non-detection and the discovery epoch, i.e. at MJD = 58880.0$\pm$1.5.

\subsubsection{UV}
Follow-up in the ultraviolet (UV), starting at +41 days, was obtained with the UltraViolet and Optical Telescope (UVOT) aboard the \textit{Neil Gehrels Swift Observatory} in the $UVW2$, $UVM2$ and $UVW1$ filters. Simultaneously, optical follow-up was obtained in the $u$, $B$ and $V$-filters. The UVOT data were retrieved from the UK Swift Data Archive\footnote{\href{https://www.swift.ac.uk/swift_portal/}{ https://www.swift.ac.uk/swift\_portal/}} and reduced using standard software distributed with \texttt{HEAsoft}\footnote{version 6.27.2, \href{https://heasarc.nasa.gov/lheasoft/}{ https://heasarc.nasa.gov/lheasoft/}}. Photometry was measured using the \texttt{FTOOLS} tasks \textit{uvotimsum} and \textit{uvotsource} with a 5\arcsec\ radius circular aperture. The host contribution has not been accounted for, but the contribution by the faint ($r \sim 23.5$ mag) host is expected to be small, see Sect.~\ref{sec:host_analysis}.

\begin{figure*}
\centering
    \includegraphics[width=0.8\textwidth]{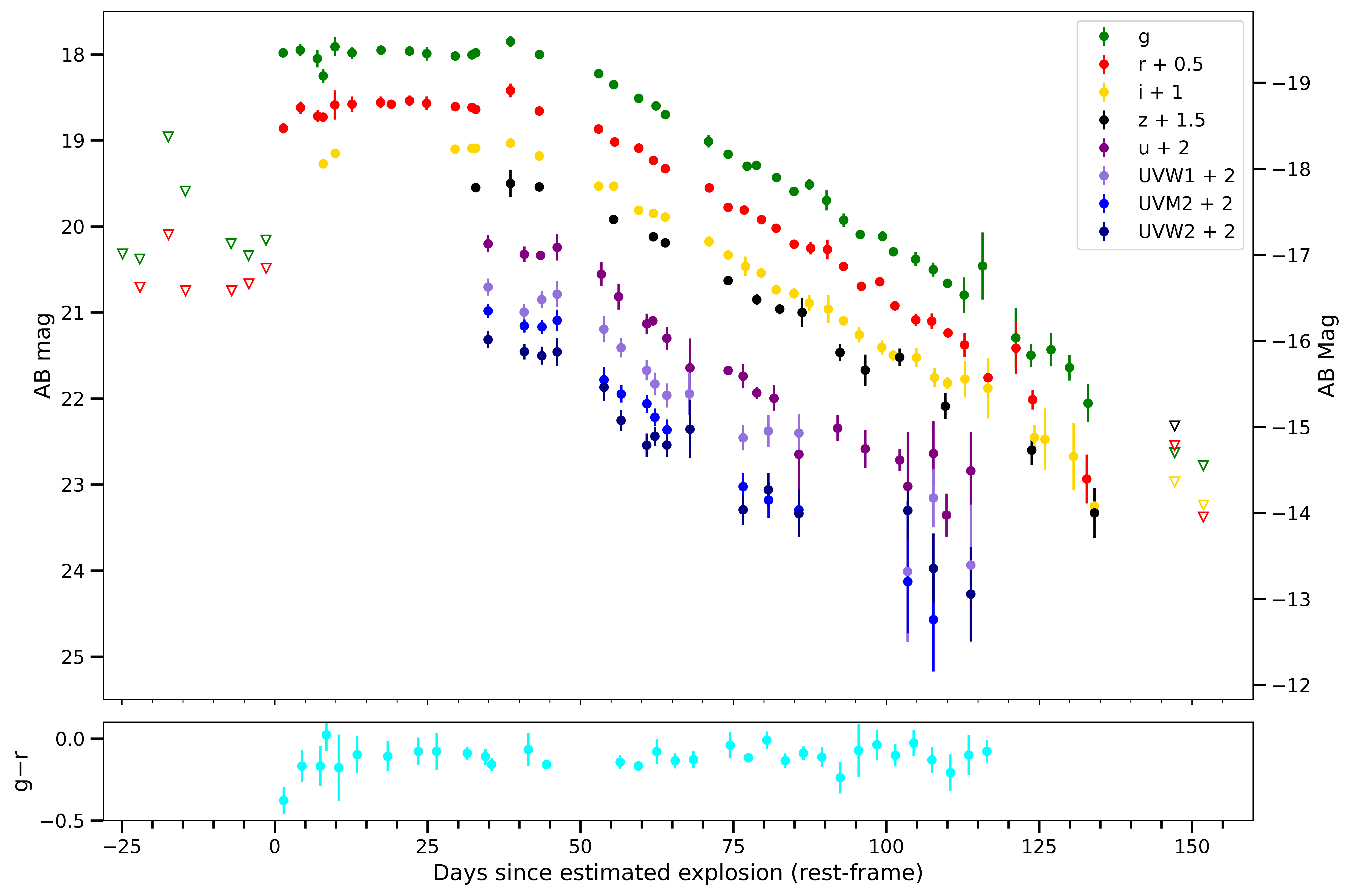}
    \caption{Lightcurves of SN\,2020bqj, corrected for MW extinction. Non-detections with 5$\sigma$ upper limits are indicated by triangles. Apparent magnitudes on the left hand axis and absolute magnitudes to the right. The photometry on the decline has been binned in three-night bins. Phase is rest-frame days since the estimated explosion epoch, which was assumed to fall halfway between latest non-detection and discovery epoch. \textit{Swift} $B$- and $V$-band data are not shown due to their large photometric uncertainties. The bottom panel shows the $g-r$ color evolution, which stays roughly constant throughout the evolution of SN\,2020bqj except for possibly the first $\sim$5 days.}
    \label{fig:lightcurve}
\end{figure*}

The UV/optical photometry in AB magnitudes are listed in Table \ref{tab:photom} and the lightcurve is shown in Fig.~\ref{fig:lightcurve}.

\subsection{Optical spectroscopy}
An initial optical spectrum was obtained 4 days after discovery, using the SEDM on the P60 telescope. Additional optical follow-up spectroscopy was obtained using SEDM, the Dual Imaging Spectrograph (DIS\footnote{\href{https://www.apo.nmsu.edu/arc35m/Instruments/DIS/}{https://www.apo.nmsu.edu/arc35m/Instruments/DIS/}}) mounted on the 3.5m telescope at the Apache Point Observatory, the Alhambra Faint Object Spectrograph and Camera (ALFOSC) on the Nordic Optical Telescope \citep[NOT;][]{djupvik2010}, SPRAT on the LT, and the Low Resolution Imaging Spectrograph \citep[LRIS;][]{oke1994} on the Keck~I telescope. The spectroscopic coverage of SN\,2020bqj presented in this paper consists of 16 spectra and extends from 4 days until 117 days after discovery in the observer frame.

\begin{figure*}
\centering
    \includegraphics[width=0.85\textwidth]{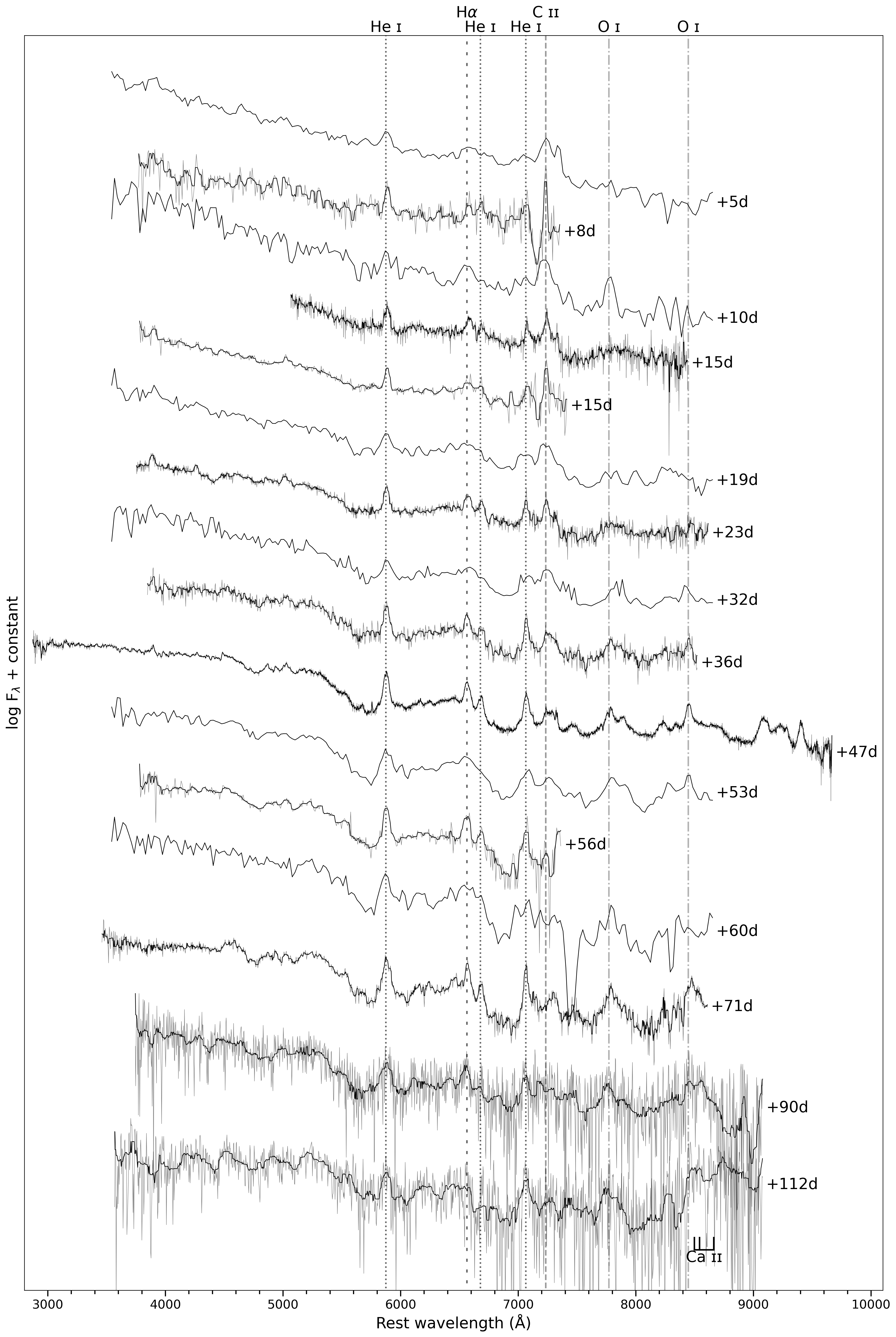}
    \caption{Spectral sequence of SN\,2020bqj, see Table~\ref{tab:spectral_log}. Phases are relative to the estimated explosion epoch, halfway between the last non-detection and the discovery epoch. Some spectra have been smoothed with a median filter for clarity. A selection of prominent emission lines are indicated.}
    \label{fig:spectra}
\end{figure*}

The spectra were reduced in a standard manner using pipelines and procedures specific for each instrument, such as \textsc{pysedm} \citep{rigault2019} for the SEDM spectra, the \textsc{PyDIS} package \citep{pydis} for APO/DIS, and \textsc{lpipe} \citep{perley2019} for Keck/LRIS. Data reduction involved bias and flat-field corrections, wavelength calibration from an arc spectrum, and flux calibration using spectrophotometric standard stars. Furthermore, all spectra were absolute flux calibrated by matching their synthetic photometry with our observed $r$-band photometry. The spectra were then corrected for MW extinction using $E(B-V)$ = 0.0194 mag and $R_V$ = 3.1. The spectral sequence is listed in Table \ref{tab:spectral_log} and shown in Fig.~\ref{fig:spectra}. All spectra will be made available via WISeREP \citep{yaron2012}.

\subsection{X-ray and radio}
The SN was observed with the X-ray telescope \citep[XRT;][]{Burrows2005a} on-board \textit{Swift} between 2020 March 9 (+35 days) and 2020 May 31 (+113 days). We analysed all data with the online-tools of the UK \textit{Swift} team\footnote{\href{https://www.swift.ac.uk/user_objects/}{https://www.swift.ac.uk/user\_objects/}} that use the methods described in \citet{Evans2007a} and \citet{Evans2009a} and the software package \texttt{HEAsoft}. The SN evaded detection in all \textit{Swift} pointings. The nominal $3\sigma$ count-rate limit varies between 0.004 and 0.03 ct/s (0.3 and 10 keV; not corrected for absorption). Using the dynamic rebinning option, the $3\sigma$ between 2020 March 9 and 2020 May 31 is $5\times10^{-4}$~ct/s (0.3-10 keV; not corrected for absorption). If we assume a power-law shaped spectrum with a photon index of 2 and a Galactic absorption of $1.67\times10^{20}~{\rm cm}^{-2}$ \citep{HI4PI2016a}, this corresponds to an absorption-corrected flux of $1.4\times10^{-14}~{\rm erg\,cm}^{-2}\,{\rm s}^{-1}$ between 0.3 and 10 keV. At the luminosity distance of SN2020bqj, this flux limits translates to a luminosity of $<2\times10^{41}~{\rm erg\,s}^{-1}$.

We obtained one epoch at radio wavelengths on 2020 March 9 (+35 days) with the Arcminute Microkelvin Imager Large Array \citep[AMI-LA;][]{zwart2008,hickish2018}. AMI-LA is a radio interferometer comprised of eight, $12.8$\,m diameter, antennas producing 28 baselines which extend from $18$ up to $110$\,m in length and operates with a $5$\,GHz bandwidth around a central frequency of $15.5$\,GHz. Initial data reduction, flagging and calibration of the phase and flux, were carried out using $\tt{reduce \_ dc}$, a customized AMI-LA data reduction software package (e.g., \citealt{perrott_2013}).  Phase calibration was conducted using short interleaved observations of J1527+3115, while 3C286 were used for absolute flux calibration. Images of the field of SN\,2020bqj were produced using CASA task CLEAN in an interactive mode, while the image rms was calculated using CASA task IMSTAT. No source was detected with a 3$\sigma$ limit of 81 $\mu$Jy. This corresponds to an upper limit of $\nu L_{\nu} \leq 1.3 \times 10^{38}~{\rm erg\,s}^{-1}$, given the luminosity distance of SN\,2020bqj.

\section{Photometric analysis} \label{sec:photom_analysis}
\subsection{Early phase lightcurve}
\begin{figure}
\centering
    \includegraphics[width=\hsize]{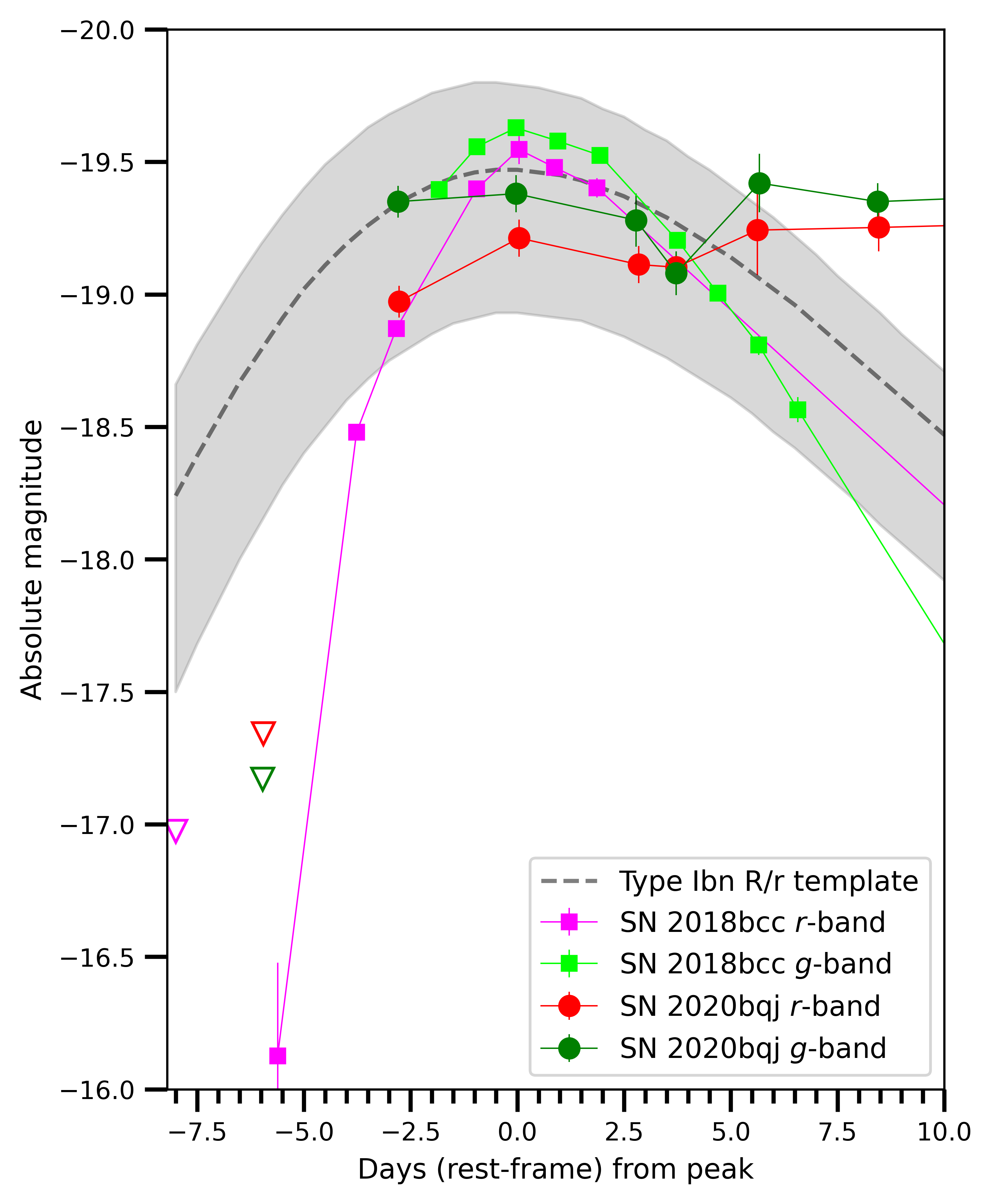}
    \caption{The $g$- and $r$-band lightcurves of SN\,2020bqj (in green and red) and the Type Ibn SN\,2018bcc (in lime and magenta) from \citet{karamehmetoglu2019} and the Type Ibn $R$/$r$-band template lightcurve from \citet{hosseinzadeh2017} at early phase. The lightcurve of SN\,2020bqj in $g$- and $r$-band in its first fifteen days is consistent with those of Type Ibn SNe, albeit relatively fast in its rise. SN\,2020bqj reached a $r$-band magnitude of the plateau consistent with the peak magnitude of the Ibn template, with a risetime constrained to < 5.7 days in rest-frame by pre-discovery upper limits. While this is fast compared to the template, it matches well with the risetime of SN\,2018bcc.} 
    \label{fig:peak_comparison}
\end{figure}

SN\,2020bqj reached its plateau magnitude in $r$-band at the second epoch (+4.2 days), with $M_r$ = $-19.23\pm0.07$, whereas in $g$-band the plateau had already been reached at the discovery epoch (+1.4 days), with $M_g$ = $-19.43\pm0.07$, see Fig.~\ref{fig:peak_comparison}.
Based on the last pre-discovery limits the risetime to the peak plateau in $r$-band is constrained to < 5.7 days in rest-frame. Such a photometric evolution is typical for a rapidly-evolving SN Ibn, as depicted by the $g$- and $r$-band lightcurves of SN\,2018bcc, a SN Ibn with a particularly well-sampled rise to peak \citep{karamehmetoglu2019}. Also shown in Fig.~\ref{fig:peak_comparison} is the Type Ibn $R$/$r$-band template lightcurve constructed by \citet{hosseinzadeh2017}, based on a sample of 18 Type Ibn SNe from the intermediate Palomar Transient Factory \citep[iPTF; ][]{kulkarni2013} and the literature. The absolute $r$-band magnitude of SN\,2020bqj of the peak plateau is consistent with the peak of the SN Ibn template. The rise to the peak plateau of SN\,2020bqj is fast compared to the template, but as noted in \citet{hosseinzadeh2017} the early phases of the SN Ibn template are biased due to the limited number of SNe Ibn in the literature with pre-peak coverage and not taking into account upper limits.
As such, a better comparison is with the lightcurve of SN\,2018bcc. The first $r$-band data point of SN\,2018bcc shown in Fig.~\ref{fig:peak_comparison} was not included in the initial study by \citet{karamehmetoglu2019}, but recovered after re-analysis of the data.

\subsection{Plateau phase and decline} \label{sec:photom_plateau}
After the initial rise, the lightcurve of SN\,2020bqj enters a plateau phase in $g$-, $r$- and $i$-band that lasts for 40 days at an $r$-band absolute magnitude ranging between $-19.1$ and $-19.3$. The plateau phase is followed by 
a linear decline lasting > 90 days with a rate of $\sim$0.04 mag day$^{-1}$. Such a post-peak evolution is in contrast with what is commonly observed in SNe Ibn, where a quick rise to peak is followed by a fast decline at a rate of $\sim$0.1 mag day$^{-1}$ \citep{hosseinzadeh2017}. 

\begin{figure}
\centering
    \includegraphics[width=\hsize]{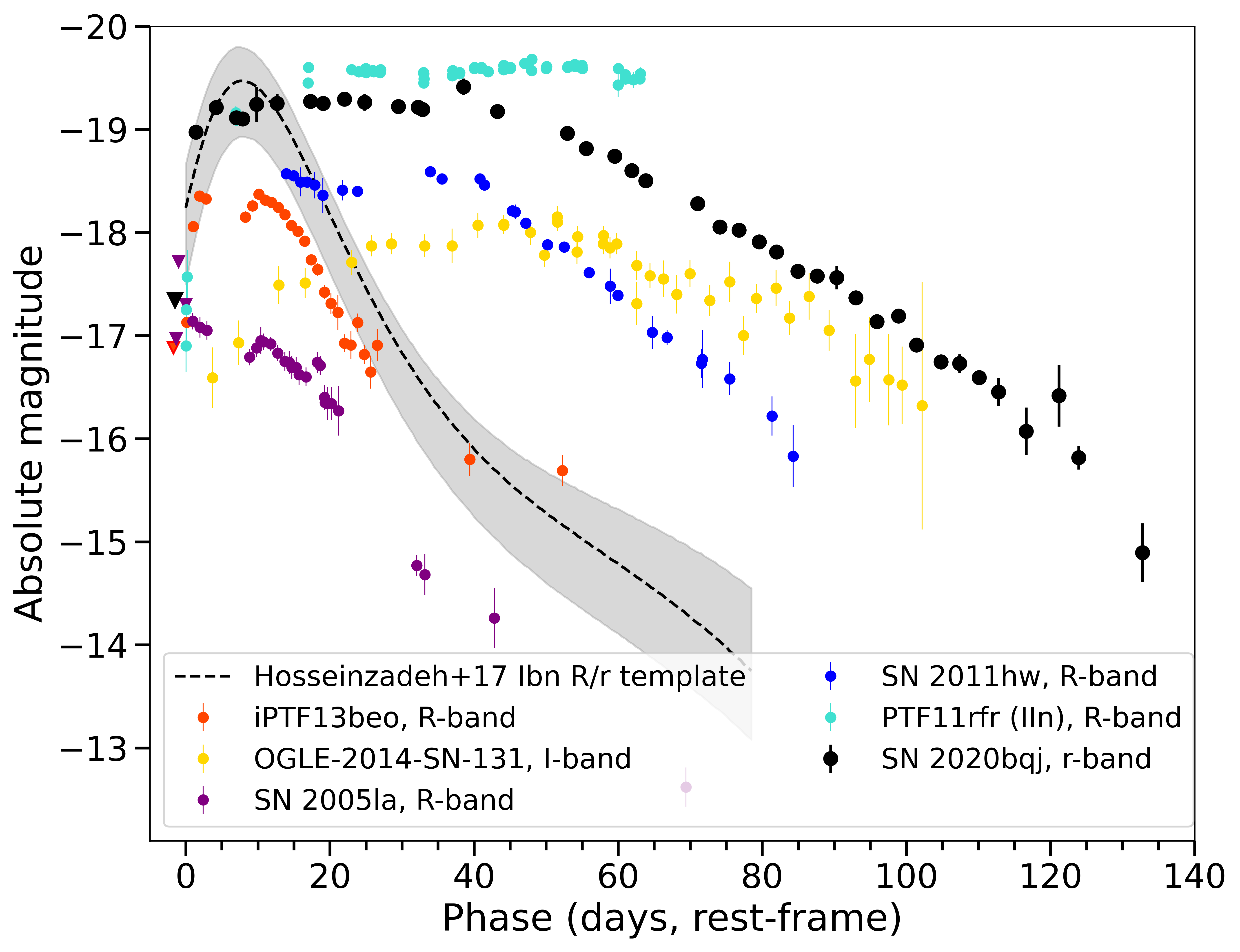}
    \caption{Type Ibn $R$/$r$-band template lightcurve from \citet{hosseinzadeh2017}, with overplotted the $r$-band lightcurve of SN\,2020bqj (in black) and the $R$/$r$/$I$-band lightcurves of the slow evolving and/or double-peaked SNe Ibn OGLE-2014-SN-131 (yellow), SN\,2011hw (blue), iPTF13beo (orange) and SN\,2005la (purple). Also shown is PTF11rfr (turquoise), a rapidly rising SN IIn with a peak plateau, but a slow decline \citep[see][]{nyholm2020}. The latest pre-discovery upper limits, where available, are indicated by triangles. While normal Type Ibn SNe have declined $\sim$4 mag 35-40 days after peak, SN\,2020bqj stayed constant in brightness during the same period. SN\,2011hw shows a similar lightcurve evolution, albeit one magnitude fainter. Phase is days in rest-frame since estimated explosion epoch for SN\,2020bqj, SN\,2011hw, iPTF13beo. The phases of SN\,2005la, OGLE-2014-SN-131 and PTF11rfr are relative to their last non-detections, although for the two SNe Ibn these observations were rather shallow. The phase of the Type Ibn template is relative to the start of the template at 8 days before peak.}
    \label{fig:template_comparison}
\end{figure}

The full $r$-band lightcurve of SN\,2020bqj and the $R/r$-band SN Ibn template are shown in Fig.~\ref{fig:template_comparison}. Where most SNe Ibn have declined $\sim$4 mag at 40 days after peak, SN\,2020bqj stayed at a similar magnitude throughout. Also plotted are the lightcurves of four other photometric outliers in the SN Ibn subclass. These SNe have irregular photometric evolution reminiscent of SN\,2020bqj, showing either a double peak or a plateau phase, and were not included in the construction of the $R/r$-band SN Ibn template. The sub-luminous OGLE-2014-SN-131 displayed a slow evolution both in the rise and the decline, peaking at an $r$-band magnitude of $M_r$ $\approx-18$ \citep{karamehmetoglu2017}. SN\,2011hw was first detected at a magnitude of $M_r \approx-18.5$, at which the SN stayed for $\sim$30 days, after which a linear decline set in with a rate of 0.055 mag day$^{-1}$ \citep{smith2012,pastorello2015IV}. No pre-discovery limits were available for SN\,2011hw, so the risetime and phase is unknown, but based on its spectral evolution \citet{pastorello2015IV} estimated the explosion epoch to have occurred 14$\pm$10 days before discovery, which we adopt in the figure. iPTF13beo reached peak magnitude 4 days after the last pre-explosion non-detection, and a second peak was observed $\sim$9 days after the first \citep{gorbikov2014}. SN\,2005la was discovered after peak, and its lightcurve showed re-brightening twice during the subsequent decline \citep{pastorello2008II}. No stringent upper limits before the discovery of SN\,2005la were available, preventing a well constrained estimate of the explosion epoch. 
The final remaining photometric outlier in the SN Ibn subclass is not shown here, since OGLE-2012-SN-006 showed a monotonic evolution similar to that of a regular SN Ibn, but at a considerable slower pace \citep{pastorello2015V}. Finally, we also show PTF11rfr, a SN IIn which rose to peak rapidly for its subclass, followed by a peak plateau and a slow decline at $\sim$ 0.01 mag day$^{-1}$ \citep{nyholm2020}.

\subsection{Pseudo-bolometric lightcurve}
A large fraction of the photometric coverage of SN\,2020bqj, from +35 days to +114 days, extends from the UV to $z$ band, covering the spectral range from $\sim$1600 to $\sim$11000~\AA. This allows us to simply integrate the spectral energy distribution (SED) using trapezoidal integration to obtain a pseudo-bolometric lightcurve that covers most of the SN flux. We interpolate the gaps in the photometric coverage in the different filters in this period using Gaussian Processes (GP) interpolation with \texttt{GPy}\footnote{\href{https://sheffieldml.github.io/GPy/}{https://sheffieldml.github.io/GPy/}}. We used multi-task GP with a radial basis function kernel to capture the smooth evolution characterized by a typical length scale, and a linear and bias kernel to capture the decline. The fits were performed in flux space with an additional white noise kernel to model the intrinsic scatter. This method will be explained in detail in Karamehmetoglu et al. (\textit{in prep}). In this way we obtain lightcurves at a nightly cadence in all filters, the integrated flux of which is used to construct the pseudo-bolometric lightcurve.

Then, we extrapolate the pseudo-bolometric lightcurve to < +35 days by applying a bolometric correction to the (interpolated) flux captured in $g$ band, where we assume that the shape of the SED of the SN does not change significantly over time. This assumption is supported by the lack of color evolution in the lightcurve (see Fig.~\ref{fig:lightcurve}), and the weak spectral evolution (see Fig.~\ref{fig:spectra} for the spectral sequence). We note that at very early times this assumption may not be correct as the SN appears bluer in the first epoch (Fig.~\ref{fig:lightcurve}).

\begin{figure}
\centering
    \includegraphics[width=\hsize]{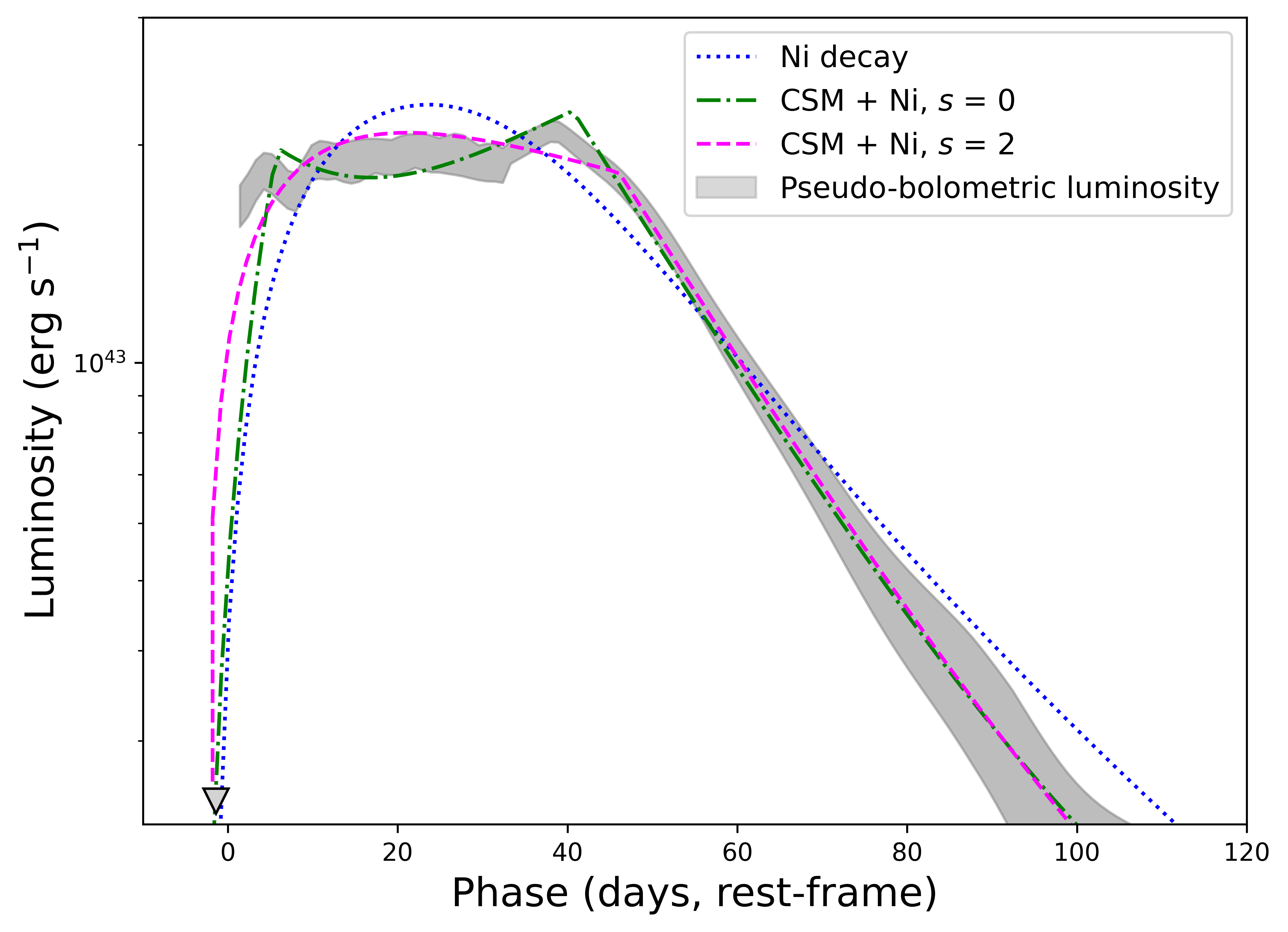}
    \caption{Pseudo-bolometric lightcurve of SN\,2020bqj based on trapezoidal integration of the broadband fluxes. Overplotted are \textsc{TigerFit} model fits (see Sect.~\ref{sec:powering}). The radioactive $^{56}$Ni decay model fit has $t_0$ set to $-2$ days before our adopted estimated explosion epoch. The hybrid CSM + Ni model fits shown assume a CSM shell ($s = 0$) or a wind profile ($s = 2$), with $t_0$ set to $-4$ and $-3$ days, respectively. The radioactive decay model does not reproduce the transitions between the rapid rise, plateau and decline phases. The hybrid models recover the shape of the lightcurve much better, in particular the Ni+CSM shell ($s=0$) model shows similar sharp transitions as the data. None of the models recover the rapid rise from the latest non-detection to discovery 2.8 rest-frame days later. Phase is relative to estimated explosion epoch.}
    \label{fig:bolometric}
\end{figure}

The resulting pseudo-bolometric lightcurve is shown in Fig.~\ref{fig:bolometric}. It is not a full bolometric lightcurve as we do not account for the SN flux in the far-UV and near-IR not covered by our photometry, but based on blackbody fits to the broadband photometry we estimate this to be < 10\% of the SN bolometric flux. Additionally, the pseudo-bolometric lightcurve does not extend to the faint end of the lightcurve past +114 days, since due to the larger uncertainties in the photometry and a lack of UV data, the faint linear tail does not meaningfully constrain the model fits discussed in Sect.~\ref{sec:powering}. The total observed radiated energy from discovery to +114 days is $1.1\pm0.1\times10^{50}$ erg. This can be considered a conservative lower limit to the total energy radiated by the SN in this period, since it does not include flux in the far-UV and IR, and the blue colors at early phases suggest a higher photospheric temperature and thus higher luminosity than was extrapolated.

\section{Spectral analysis} \label{sec:spectra_analysis}
\subsection{Line identification}
Throughout the evolution of SN\,2020bqj the spectra of the SN are dominated by intermediate-width emission lines ($\sim2000$ km s$^{-1}$, further discussed in Sect.~\ref{sec:line_evo}). The initial low resolution SEDM spectrum obtained at +5 days (as with the lightcurve, phase is rest-frame days since estimated explosion epoch) did not allow for a secure classification of the object, but based on subsequent spectra the prominent emission lines in the earliest spectrum can be identified as He\,{\sc i} $\lambda\lambda$5876, 7065, and C\,{\sc ii} $\lambda$ 7231--7236, see Fig.~\ref{fig:spectra}. In the noisier blue end of the spectrum also C\,{\sc ii} $\lambda\lambda$3920,4267 can be identified, where the bluest line is likely merged with He\,{\sc i} $\lambda$3888.
Subsequent spectra continue to show the same strong He\,{\sc i} emission lines, as well as C\,{\sc ii}, most prominently at 7231--7236~\AA. 

\begin{figure*}
\centering
    \includegraphics[width=\textwidth]{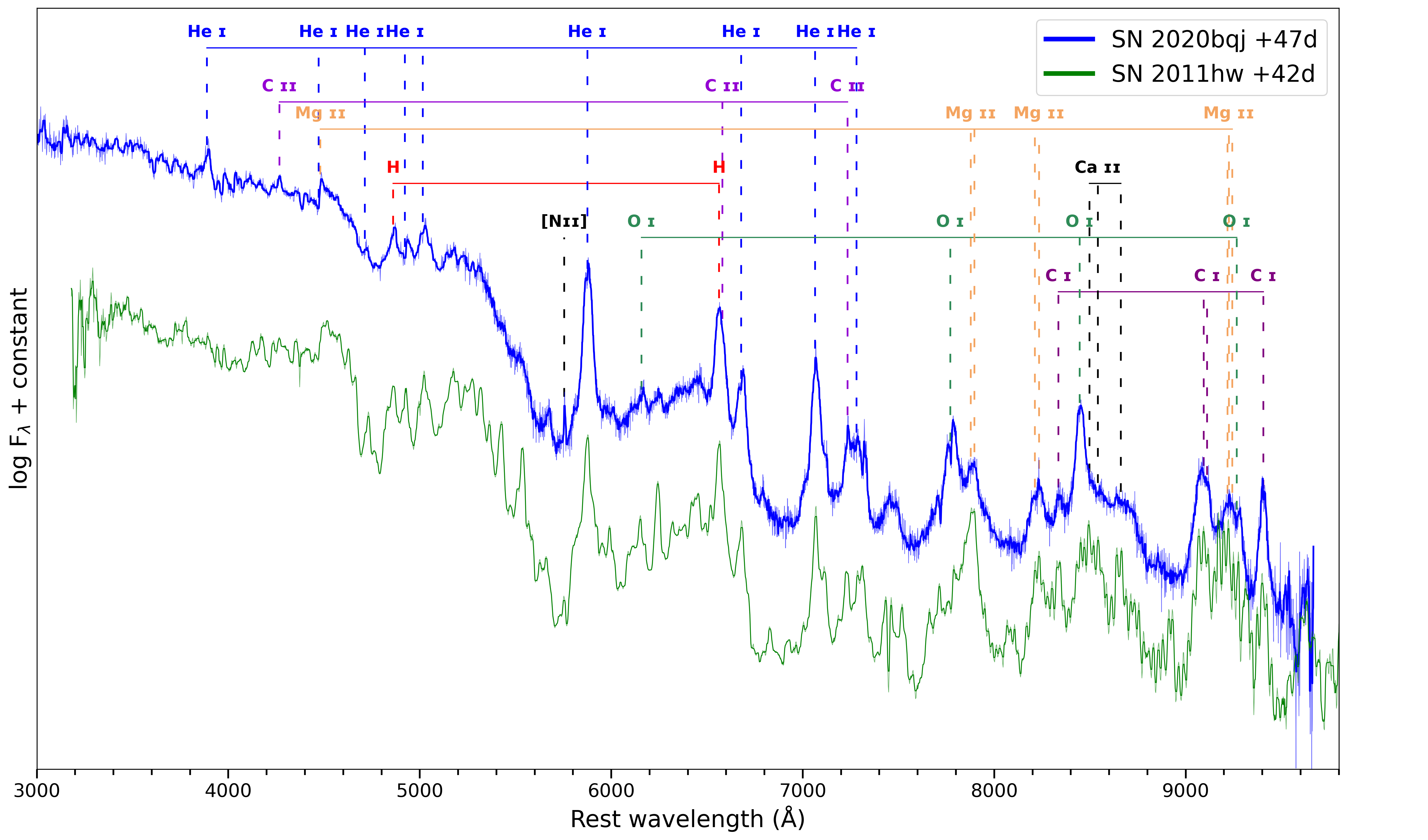}
    \caption{High signal-to-noise spectrum of SN\,2020bqj obtained with Keck at +47 days after estimated explosion epoch, with all emission lines discussed in the text identified. Also shown is a TNG/DOLORES spectrum of the peculiar SN\,2011hw obtained +42 days after estimated explosion epoch \citep{pastorello2015IV}. SN\,2011hw is a SN Ibn with a photometric and spectroscopic evolution that is very similar to SN\,2020bqj, albeit somewhat faster. For example, the Ca\,{\sc ii} triplet is already visible in SN\,2011hw, while in SN\,2020bqj this feature only shows up from +71 days onwards (Fig.\ref{fig:spectra}).}
    \label{fig:line_identification}
\end{figure*}

The high signal-to-noise (SNR) spectrum obtained at +47 days with Keck allows for secure line identifications and line velocity measurements, see Fig.~\ref{fig:line_identification}. FWHM line velocities were measured by fitting a Gaussian line profile, with errors dominated by the resolution of our low-resolution spectrographs. In addition to He\,{\sc i} $\lambda\lambda$5876, 7065, the Keck spectrum shows clear He\,{\sc i} in emission at $\lambda\lambda$3888, 5016, 6678 and 7281, with line velocities ranging from $v_{\mathrm{FWHM}} = 1500\pm130$ km s$^{-1}$ for He\,{\sc i} $\lambda$7281 to $v_{\mathrm{FWHM}} = 2680\pm130$ km s$^{-1}$ for He\,{\sc i} $\lambda$5876.
H$\alpha$ and H$\beta$ are visible, with H$\alpha$ possibly merged with faint C\,{\sc ii} $\lambda$6580 (see Sect.~\ref{sec:line_evo}). Assuming H$\alpha$ dominates this emission feature, the mean Balmer velocity is $v_{\mathrm{FWHM}} = 2370\pm140$ km s$^{-1}$, consistent with He\,{\sc i} $\lambda$5876. Although weak at this epoch, C\,{\sc ii} $\lambda$4267 and C\,{\sc ii} $\lambda$7231--7236 are still discernible, where the latter line is blended with He\,{\sc i} $\lambda$7281. By fitting two Gaussians simultaneously, we measure a velocity of $v_{\mathrm{FWHM}} = 1870\pm140$ km s$^{-1}$ for C\,{\sc ii} $\lambda$7231--7236. C\,{\sc i} is detected at $\lambda\lambda$8335, 9094--9111 and 9405, with a velocity of $v_{\mathrm{FWHM}} = 1380\pm160$ km s$^{-1}$.
O\,{\sc i} at 7772--7775~\AA\ and 8446~\AA\ are present, the latter of which has a velocity of $v_{\mathrm{FWHM}} = 2030\pm160$ km s$^{-1}$. Mg\,{\sc ii} at $\lambda\lambda$7877--7896 and $\lambda\lambda$8213--8234 are visible, as well as at $\lambda\lambda$9218--9244 potentially merged with O\,{\sc i} $\lambda$9266. The steep pseudo-continuum bluewards of $\sim$5700~\AA\ is typical for SNe Ibn post-peak, and is attributed to the blending of a forest of Fe lines \citep{smith2012,Stritzinger2012,pastorello2015IV} Finally, the position of the Ca\,{\sc ii} triplet is indicated, which is not observed yet at this epoch, but does become prominent at later times, see Fig.~\ref{fig:spectra}. 

Also shown in Fig.~\ref{fig:line_identification} is a TNG/DOLORES spectrum of SN\,2011hw \citep{pastorello2015IV}, a peculiar SN Ibn with a photometric evolution similar to that of SN\,2020bqj, see Sect.~\ref{sec:photom_plateau}. This spectrum was obtained at a similar epoch as the Keck spectrum of SN\,2020bqj, just after the decline has set in, and is remarkably similar. All aforementioned lines were also observed in SN\,2011hw, although O\,{\sc i} is weaker, Mg\,{\sc ii} more prominent, and Ca\,{\sc ii} is already visible (in SN\,2020bqj Ca\,{\sc ii} is not observed until at +71 days, see Fig.~\ref{fig:spectra}). The intermediate-width He\,{\sc i} lines in SN\,2011hw have a similar velocity ($\sim$1900 km s$^{-1}$, \citealt{smith2012}) as SN\,2020bqj. The spectra comparison also allows us to identify the weak narrow (unresolved, $\lesssim300$ km s$^{-1}$) feature blue-wards of He\,{\sc i} $\lambda$5876 as [N\,{\sc ii}] $\lambda$5754, which was also detected in SN\,2011hw \citep{pastorello2015IV}. It is worth noting that C\,{\sc i} and C\,{\sc ii} are also visible in the SN\,2011hw spectrum, although this was not reported in the original studies \citep{smith2012,pastorello2015IV}. In particular at early epochs the spectra of SN\,2011h show prominent C\,{\sc ii}, which is further discussed below.

\subsection{Spectral line evolution} \label{sec:line_evo}
\begin{figure}
\centering
    \includegraphics[width=\hsize]{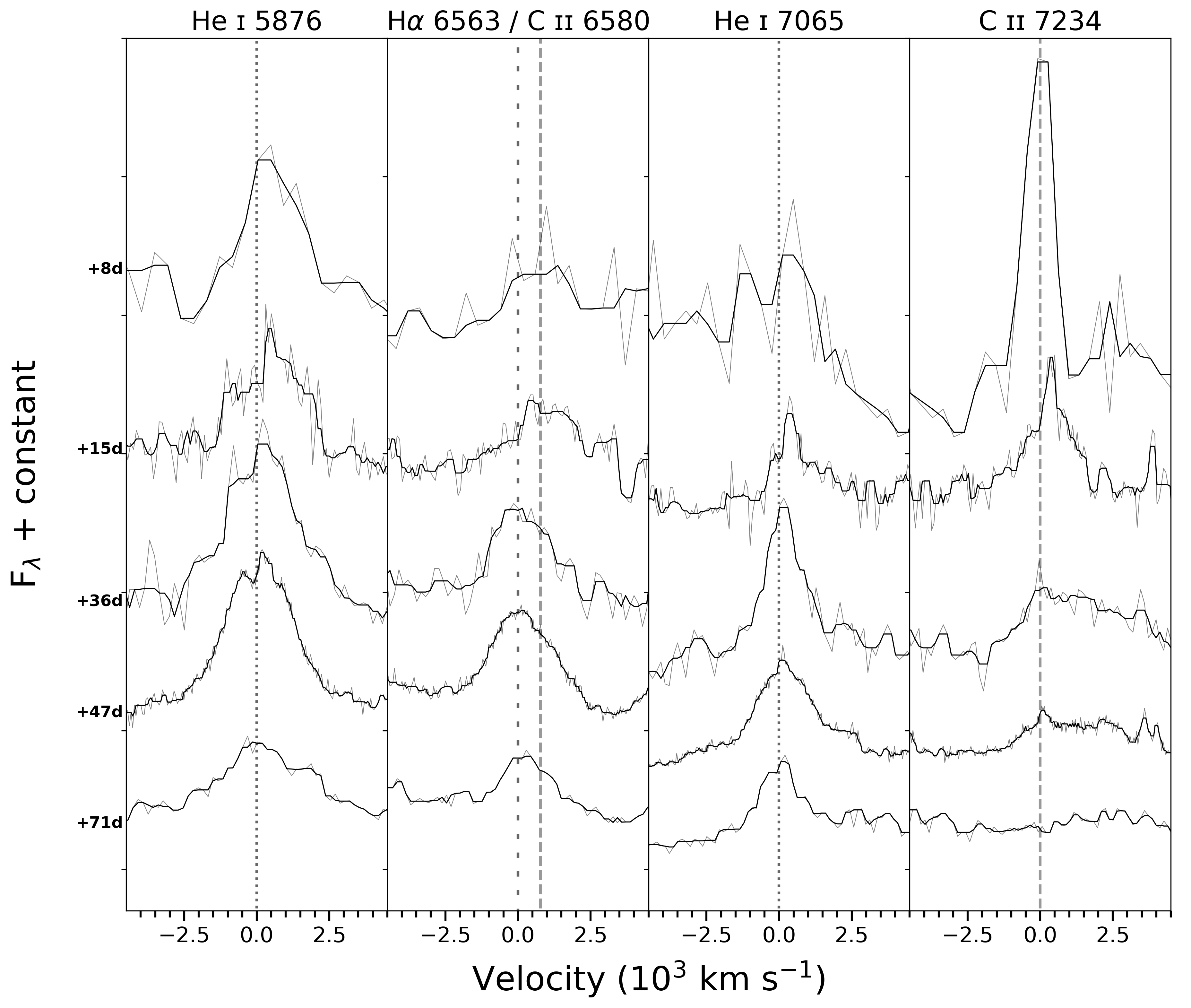}
    \caption{Line evolution of prominent emission lines in the spectra of SN\,2020bqj. From left to right: He\,{\sc i} $\lambda$5876, H$\alpha$ merged with C\,{\sc ii} $\lambda$6580, He\,{\sc i} $\lambda$7065 and C\,{\sc ii} $\lambda$7231--7236.
    }
    \label{fig:line_evolution}
\end{figure}

Figure~\ref{fig:line_evolution} shows the evolution of some of the most prominent emission lines in the spectra of SN\,2020bqj. In summary, throughout the spectral sequence He\,{\sc i} $\lambda$5876 and He\,{\sc i} $\lambda$7065 remain prominent, whereas the C\,{\sc ii} $\lambda$7231--7236 emission line decreases in strength over time and appears to have completely disappeared at +71 days. The line feature at 6580~\AA\ at early epochs likely consists of H$\alpha$ and C\,{\sc ii} $\lambda\lambda$6578--6582, but by +47 days in the high SNR Keck spectrum the emission feature is well described by a single Gaussian centered on H$\alpha$, implying C\,{\sc ii} $\lambda\lambda$6578--6582 has faded away.

We quantified the evolution in line strength and velocity of the He\,{\sc i} $\lambda\lambda$5876, 7065 and C\,{\sc ii} $\lambda$7231--7236 lines by fitting a Gaussian line profile. A flat continuum around the line was assumed and derived by fitting the region surrounding the emission line, while excluding the emission line itself. We then fit a Gaussian to the line after subtracting the straight-line continuum. We obtained FWHM velocities and pseudo-Equivalent Widths (pEWs) from our Gaussian fits using standard formulae. We corrected the Gaussian $\sigma$ for instrumental resolution of each spectrum at the position of the line.
All fits were visually inspected and any poor fits were removed. This was usually due to poor SNR at the line position from a noisy spectrum. In addition, the fits to the P60 spectra, which have very low instrumental resolution, are not included in our analysis.
The statistical uncertainties of the fits were derived from the fitting errors. However, these are only a minor source of uncertainty in fitting emission lines. The dominant source of uncertainty arises from the inability to distinguish the continuum from the line. As an estimate for this source of uncertainty, we used the FWHM of the instrumental resolution at the position of the respective line, added in quad with the fitting errors. The reason we used the instrumental resolution as a proxy for this uncertainty is because the resolution acts as a natural measure of line blending, which is one of the main reasons why separating the continuum from the line is so difficult. 

\begin{figure}
    \centering
    \includegraphics[width=\hsize]{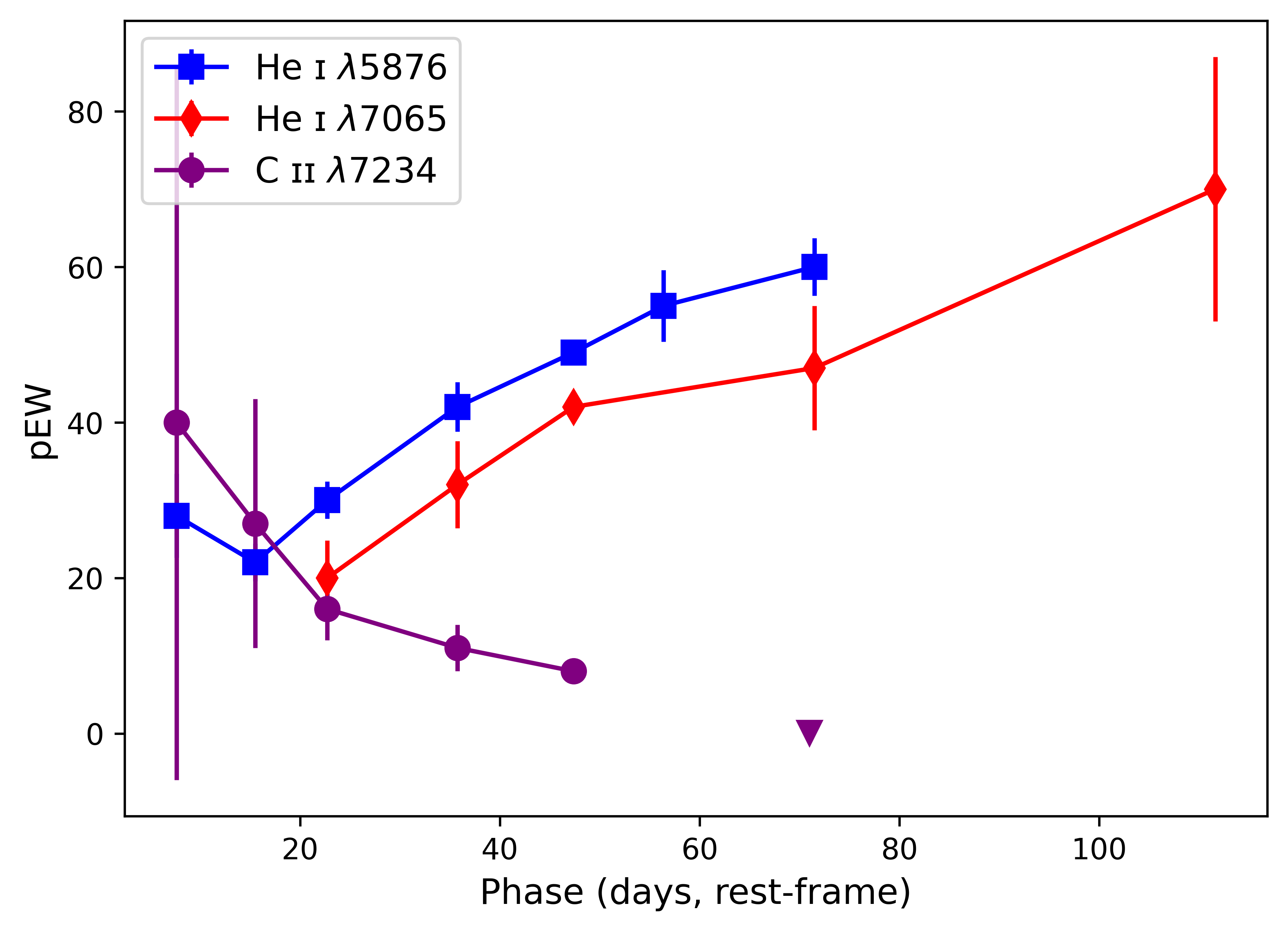}
    \caption{
    Evolution of the pEW for the He\,{\sc i} $\lambda\lambda$5876, 7065 and C\,{\sc ii} $\lambda$7231--7236 emission lines. P60 spectra have not been analysed due to their lower resolution. Fits to low SNR lines are excluded, especially for the latter spectra.}
    \label{fig:pewandvfwhm}
\end{figure}

The resulting pEW evolution is shown in Fig.~\ref{fig:pewandvfwhm}. Table~\ref{tab:pew_fwhm} lists the pEW and $v_{\mathrm{FWHM}}$, with the low resolution SEDM spectra and poor line fits omitted.

\begin{table*}
\caption{\label{tab:pew_fwhm}Line velocities and pseudo equivalent width of the He\,{\sc i} $\lambda\lambda$5876, 7065 and C\,{\sc ii} $\lambda$7231--7236 emission lines.}
\centering
\begin{tabular}{ccccccccc} \hline \noalign{\vskip 5pt}
UT Date & MJD &Phase&\multicolumn{2}{c}{He\,{\sc i} $\lambda$5876}& \multicolumn{2}{c}{He\,{\sc i} $\lambda$7065} &  \multicolumn{2}{c}{C\,{\sc ii} $\lambda$7231--7236}\\ \hline \noalign{\vskip 5pt}
&&&pEW&$v_{\mathrm{FWHM}}$ &pEW&$v_{\mathrm{FWHM}}$&pEW&$v_{\mathrm{FWHM}}$ \\ 
&&(days)&&(km s$^{-1}$)&&(km s$^{-1}$)&&(km s$^{-1}$)\\ \hline \noalign{\vskip 5pt}
2020-02-09.1 &  58888.1  &+8     &28 $\pm$ 5.4 &1900 $\pm$ 400    &             &               &40 $\pm$ 46&540 $\pm$ 370\\
2020-02-17.5 &  58896.5  &+15    &22 $\pm$ 2.4 &1990 $\pm$ 370    &             &               &27 $\pm$ 16&790 $\pm$ 370\\
2020-02-25.2 &  58904.2  &+23    &30 $\pm$ 2.4 &2360 $\pm$ 360    &20 $\pm$ 4.8 &1220 $\pm$ 360 &16 $\pm$ 4 &1310 $\pm$ 360\\
2020-03-10.1 &  58918.1  &+36    &42 $\pm$ 3.2 &2610 $\pm$ 360    &32 $\pm$ 5.6 &1390 $\pm$ 360 &11 $\pm$ 3 &1670 $\pm$ 380\\
2020-03-22.5 &  58930.5  &+47    &49 $\pm$ 0.9 &2680 $\pm$ 130    &42 $\pm$ 1.1 &1500 $\pm$ 130 & 8 $\pm$ 1 &1870 $\pm$ 140\\
2020-04-01.1 &  58940.1  &+56    &55 $\pm$ 4.6 &2760 $\pm$ 370    &             &               &           &\\
2020-04-17.2 &  58956.2  &+71    &60 $\pm$ 3.7 &3240 $\pm$ 360    &47 $\pm$ 8   &1440 $\pm$ 360 &<0.05      &\\
2020-05-30.0 &  58999.0  &+112   &             &                  &70 $\pm$ 17&1740 $\pm$ 370 &           & \\
\hline
\end{tabular}
\end{table*}

The first trend in the He\,{\sc i} lines that is worth noting is the evolution of the relative pEW of He\,{\sc i} $\lambda$5876 and He\,{\sc i} $\lambda$7065. While He\,{\sc i} $\lambda$5876 initially has twice the pEW of He\,{\sc i} $\lambda$7065, both become stronger with time and this difference seems to lessen. In addition, He\,{\sc i} $\lambda$7065 is weak or possibly absent in our first spectrum at +8 days, while He\,{\sc i} $\lambda$5876 is clearly present. An evolution of He\,{\sc i} $\lambda$7065 becoming relatively stronger with time and growing to match He\,{\sc i} $\lambda$5876 is also observed in other SNe Ibn \citep[e.g.][]{karamehmetoglu2019} and is an indication of very high electron densities in the post-shock shell \citep{almog1989,smith2012}. Secondly, the FWHM velocities of He\,{\sc i} $\lambda$7065 are significantly slower than those of He\,{\sc i} $\lambda$5876.

The pEW and velocity evolution of the C\,{\sc ii} $\lambda$7231--7236 line doublet is markedly different. Initially the line is possibly even stronger than He\,{\sc i} $\lambda$5876, which despite the large uncertainty of the pEW measurement (the line is barely resolved in the LT/SPRAT spectrum) is also apparent in the spectral sequence shown in Fig.~\ref{fig:line_evolution}. Subsequently, the line strength of C\,{\sc ii} decreases and by the time of the NOT spectrum at +71 days, it is impossible to say whether the C\,{\sc ii} $\lambda$7231--7236 line is still present or not with a pEW and line velocity consistent with zero.
As discussed previously, a similar behavior is observed in all C lines we observe. Additionally, the line velocity of C\,{\sc ii} $\lambda$7231--7236 is consistently lower than that of He\,{\sc i} $\lambda$5876 by $\sim$1700 km s$^{-1}$, which is especially noticeable in the early spectra.

\begin{figure*}
\centering
    \includegraphics[width=\textwidth]{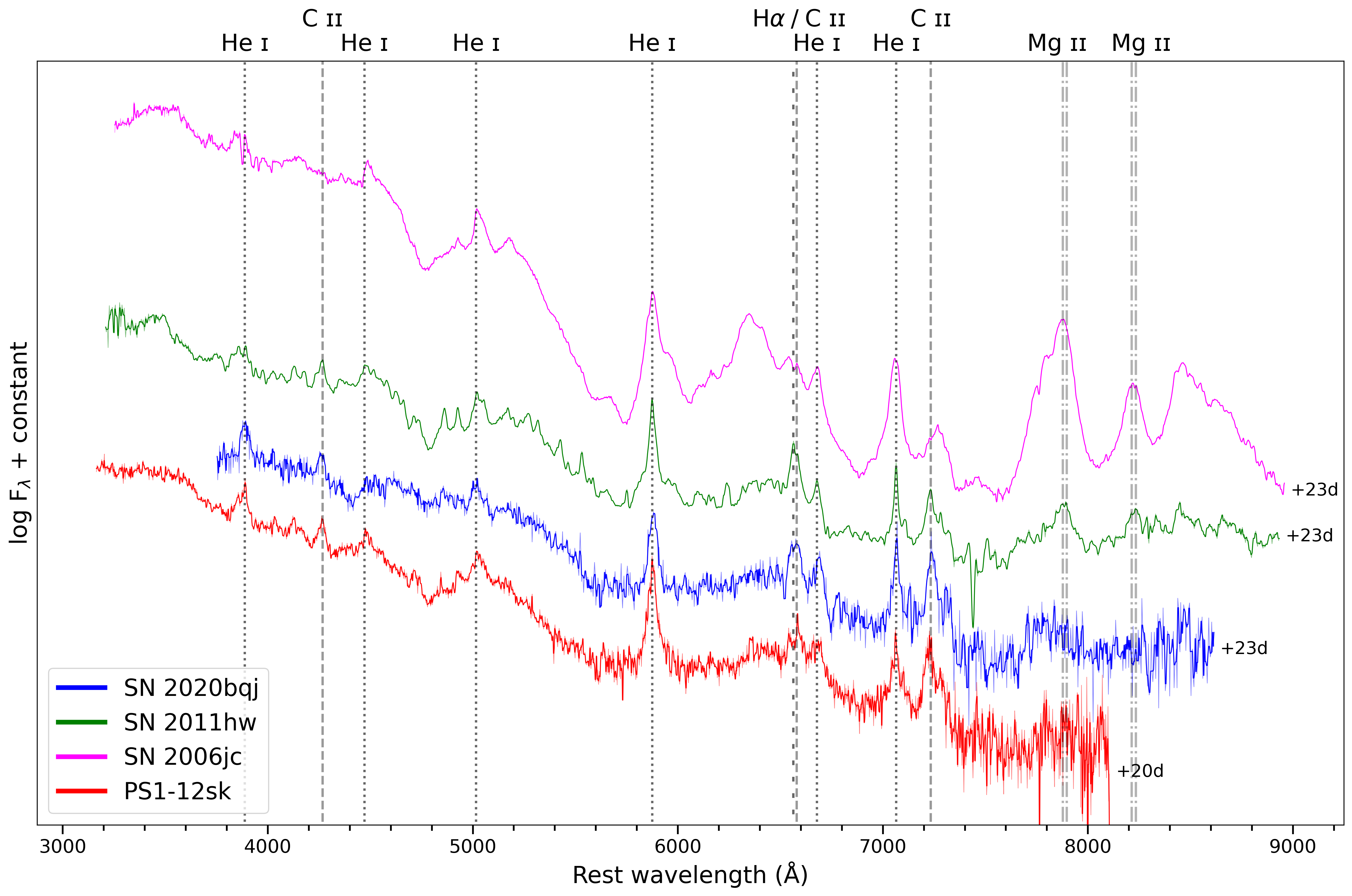}
    \caption{Spectrum of SN\,2020bqj at +23 days, together with spectra of the prototypical SN Ibn 2006jc and the peculiar SNe Ibn SN\,2011hw and PS1-12sk, at similar epochs since estimated explosion. SN\,2020bqj, SN\,2011hw and PS1-12sk show similar features across the spectrum, including prominent C\,{\sc ii} lines, which are not observed in SN\,2006jc. The C\,{\sc ii} features in the spectrum of PS1-12sk have disappeared at +26 days \citep[see][]{sanders2013}, whereas for SN\,2020bqj and SN\,2011hw they remained visible for $\sim$50 days.}
    \label{fig:12sk_comparison}
\end{figure*}

C\,{\sc ii} emission lines are not commonly observed in SNe Ibn, but as discussed earlier they featured in early spectra of SN\,2011hw. Additionally, an early spectrum of PS1-12sk \citep{sanders2013} showed intermediate-width ($\sim$2000 km s$^{-1}$) C\,{\sc ii}, and broad C\,{\sc ii} (possibly merged with He\,{\sc i} $\lambda$7281) was observed in LSQ13ccw \citep{pastorello2015VI}. Interestingly, both SN\,2011hw and PS1-12sk are considered outliers in the SN Ibn subclass based on photometric evolution and host galaxy properties, respectively, and both showed H$\alpha$ as well. In Fig.~\ref{fig:12sk_comparison} we show early spectra of SN\,2011hw and PS1-12sk compared to spectra of SN\,2020bqj at similar epochs, as well as the prototypical Type Ibn SN\,2006jc \citep{foley2007}. Phases are relative to estimated explosion epochs. As with SN\,2011hw (Sect.~\ref{sec:photom_plateau}), the explosion epochs of SN\,2006jc and PS1-12sk are poorly constrained. For PS1-12sk we adopt the estimated explosion epoch of MJD = 55992$\pm$5 from \citet{hosseinzadeh2017}, and for SN\,2006jc we adopt as explosion epoch MJD = 54004, halfway between last non-detection and the estimated peak epoch of MJD = 54008 from \citet{hosseinzadeh2017}.

At $\sim$22 days past explosion, SN\,2020bqj, SN\,2011hw and PS1-12sk all show very similar features across the spectrum, featuring prominent intermediate-width C\,{\sc ii} and H$\alpha$ in addition to the canonical He lines. As with SN\,2020bqj, the C\,{\sc ii} emission lines fade away with time, although at different rates. A spectrum obtained of PS1-12sk at +26 days showed C\,{\sc ii} $\lambda$7231--7236 as an absorption feature instead \citep{sanders2013}, while the same feature was weak but still observed at $\sim$+45 days in both SN\,2020bqj and SN\,2011hw, see Fig.~\ref{fig:line_identification}. No carbon lines were ever reported for SN\,2006jc. \citet{sanders2013} attributed the initial presence and fast decline of C\,{\sc ii} to an effective temperature dependence, as is observed in Type Ia SNe \citep{parrent2011}. Such a temperature decrease is seemingly in conflict with the lack of color evolution of SN\,2020bqj (Fig.~\ref{fig:lightcurve}). This may suggest that the lack of color evolution is a result of (ongoing) CSM interaction features dominating the SN spectrum, such as the pseudo-continuum due to Fe lines in $g$-band. The only Balmer line detected in SN\,2006jc was a weak H$\alpha$ emission line that increased slightly in line strength throughout its evolution \citep{foley2007}. SN\,2020bqj, SN\,2011hw and PS1-12sk also all show H$\alpha$ increasing with time \citep{smith2012,sanders2013}, but the line is much more prominent in SN\,2020bqj and SN\,2011hw, particular at late epochs, see Fig.~\ref{fig:line_identification}.

Finally, studying the evolution of the different velocity components in the emission lines of interacting SNe provides a method to constrain the properties of the CSM and origin of the line-emitting regions. Intermediate-width ($\sim$few$\times10^3$ km s$^{-1}$) and broad (up to $\sim10^4$ km s$^{-1}$) emission lines in SNe Ibn are thought to originate from either the shock front between SN ejecta and CSM, or from the freely expanding SN ejecta, respectively \citep{pastorello2016}. 

The velocities of the intermediate-width He emission lines of SN 2020bqj evolve only slowly over time (Table~\ref{tab:pew_fwhm}), and show no evidence of a broad P-Cygni profile. Based on this, we interpret the intermediate-width emission lines of $\sim2500$ km s$^{-1}$ to be that of the shocked gas, rather than the SN ejecta. Narrow ($10^2 - 10^3$ km s$^{-1}$) velocity components superimposed on the strong He lines, such as observed in SN\,2011hw \citep{smith2012}, likely trace unshocked slow-moving CSM. The velocity of the unperturbed CSM can be determined from the narrow emission line widths, or inferred from the blueshifted absorption minimum of a narrow P-Cygni profile, if present \citep{pastorello2016}. SN\,2020bqj shows some evidence of narrow emission features superimposed on broad He lines, but they are weak and unresolved. However, weak blue-shifted absorption superimposed on the strong emission lines is observed in for example He\,{\sc i} $\lambda$5876 (Fig.~\ref{fig:line_evolution}). We measure from a selection of He lines in the +47 days Keck spectrum an absorption velocity of 230 $\pm$ 130 km s$^{-1}$, relative to the intermediate-width emission line centers. Based on the analysis of similar features in SN Ibn LSQ13ddu \citep{clark2020} in low and high resolution spectra, this velocity can be considered an upper limit to the velocity of the unperturbed CSM around SN\,2020bqj due to our lack of spectra with high spectral resolution.

\section{Host galaxy} \label{sec:host_analysis}
To investigate the properties of the host galaxy of SN\,2020bqj we modelled its SED based on archival data retrieved from the Legacy Surveys \citep{dey2019}. We measured the brightness of the host using the aperture photometry tool presented in \citet{Schulze2018a} that is based on Source Extractor version 2.19.5 \citep{Bertin1996a}. To calibrate the instrumental magnitudes, we measured the brightness of stars from SDSS in the same way. The observed AB magnitudes (not corrected for extinction) of the host obtained this way are $g = 23.38\pm0.18$, $r = 23.46\pm0.40$ and $z > 22.2$. We modelled the SED with the software package \texttt{prospector} version 0.3 \citep{Leja2017a}. Prospector uses the \texttt{Flexible Stellar Population Synthesis} (FSPS) code \citep{Conroy2009a} to generate the underlying physical model and \texttt{python-fsps} \citep{ForemanMackey2014a} to interface with FSPS in python. The FSPS code also accounts for the contribution from the diffuse gas (e.g., \ion{H}{II} regions) based on the Cloudy models from \citet{Byler2017a}. Furthermore, we assumed a Chabrier initial mass function \citep{Chabrier2003a} and approximated the star formation history (SFH) by a linearly increasing SFH at early times followed by an exponential decline at late times (functional form $t \times \exp\left(-t/\tau\right)$). The model was attenuated with the \citet{Calzetti2000a} model. Finally, we use the dynamic nested sampling package \texttt{dynesty} \citep{Speagle2020a} to sample the posterior probability function.

\begin{figure}
    \centering
    \includegraphics[width=\hsize]{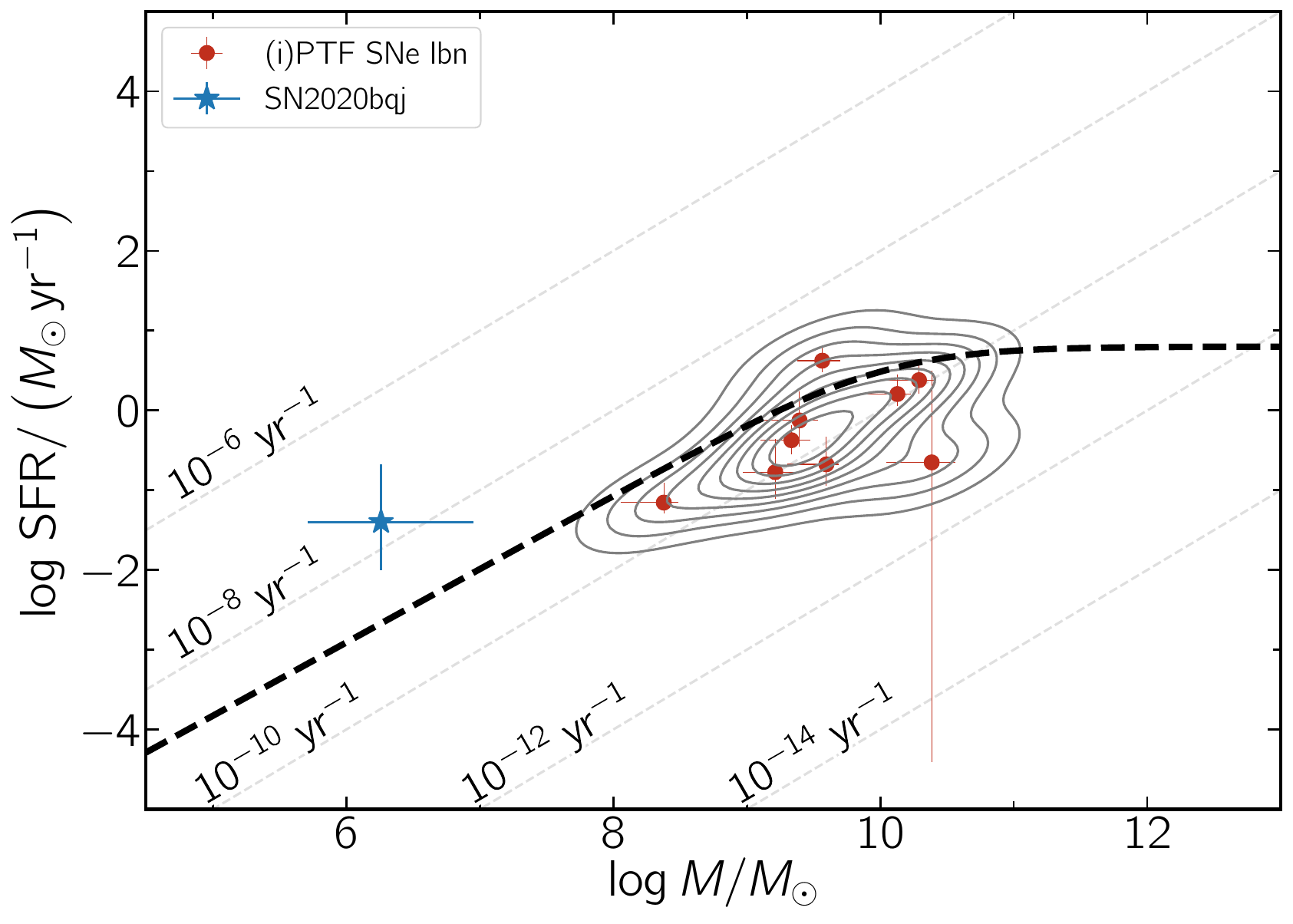}
    \caption{The host galaxy of SN\,2020bqj in the mass-SFR plane. The hosts of SNE Ibn from the iPTF survey \citep{schulze2020} have masses between $10^8$ and $10^{10}~M_\odot$ and lie in the region of the galaxy main sequence, a fundamental correlation between SFR and mass of star-forming galaxies (thick dashed line; \citealt{Lee2015a}). To guide the eye, contours from 10\% to 90\% based on the iPTF sample are overlaid. The mass of SN\,2020bqj's host stands out from this comparison sample. Its mass is more than 2 dex lower than the least massive host of that sample, albeit the errors are large. The grey-dashed lines display lines of constant specific star-formation rate (SFR normalised by stellar mass). } 
    \label{fig:ptf_hosts}
\end{figure}

The derived mass and star-formation rate of the galaxy are $\log (M/M_\odot) = 6.26^{+0.69}_{-0.55}$ and ${\rm SFR} = 0.04^{+0.17}_{-0.03} M_\odot\,{\rm yr}^{-1}$, respectively. The values represent the median of the posterior probability function and $1\sigma$ errors. The best-fit SED has a reduced $\chi^2$ of 3.5 for three filters. Based on the derived parameters, the host of SN\,2020bqj is an outlier in the SN Ibn subclass, with in particular a low galaxy mass. This is demonstrated in Fig. \ref{fig:ptf_hosts}, where the derived properties of the host of SN\,2020bqj are shown together with those of a sample of SN Ibn hosts from iPTF, where mass and SFR was derived in a consistent way \citep{schulze2020}. The host galaxy masses of the nine SNe Ibn observed in iPTF ranged between $10^8$ to $10^{10}$ M$_{\odot}$, which puts the mass of the host of SN\,2020bqj two orders of magnitude lower than the least massive SN Ibn host in iPTF. In terms of total SFR, the host of SN\,2020bqj is low but within uncertainties consistent with the lower end of the SFR distribution of the iPTF sample. As a consequence of its very low mass, the host of SN\,2020bqj falls above the galaxy main sequence, a fundamental correlation between SFR and mass of star-forming galaxy \citep{Lee2015a}. The specific SFR of the host, SFR normalised by stellar mass, is high in comparison with the iPTF host sample, and in the range of starburst galaxies. Finally, at a distance of 297 Mpc, the host has an absolute magnitude of $M_r = -14.35$, which is considerably fainter than the sample discussed in \citet{pastorello2015V} of actively starforming spirals that typically host SNe Ibn. 

In conclusion, SN\,2020bqj was hosted by an unusually faint and small galaxy for a SN Ibn host. Here it must be noted that the SED model of the host is based on sparse archival SN-free imaging. In order to further constrain the hosts properties, deep imaging and spectroscopy are required, to be obtained after the SN has faded. Such follow-up would also allow for the determination of the physical size of the host, and then a direct comparison of its SFR density with those of the sample of SN Ibn hosts presented by \citet{hosseinzadeh2019}. Constraints on the host galaxy metallicity would also be useful, as a low metallicity environment has implications on the progenitor, and its mass-loss rate, of SN\,2020bqj. Interestingly, the prominent outlier in the study by \citet{hosseinzadeh2019} in terms of host galaxy properties, PS1-12sk, was a SN Ibn discovered in an elliptical cluster galaxy at a site of extremely low local SFR density. As discussed in Sect.~\ref{sec:spectra_analysis} and shown in Fig.~\ref{fig:12sk_comparison}, PS1-12sk is spectroscopically very similar to SN\,2020bqj.

\section{Powering mechanism} \label{sec:powering}
Normal Type Ib/c SE SNe have lightcurves that are well fitted by $^{56}$Ni decay powered models \citep[e.g.,][]{prentice2016}. The lightcurves of SNe Ibn usually call for a different and/or additional power source, because the amount of $^{56}$Ni required to explain their peak luminosity can not be reconciled with the steep decline rate of SNe Ibn. Although the decline rate of SN\,2020bqj is much slower, the sharp transitions between rise, plateau and decline in the lightcurve of SN\,2020bqj do not match with the monotonic evolution of a radioactive decay dominated lightcurve. In the spectra of SN\,2020bqj intermediate-width emission lines are visible throughout the full lightcurve evolution (Fig.~\ref{fig:spectra}), so it is only natural to assume CSM interaction plays an important role in the powering of the peculiar lightcurve of SN\,2020bqj.

We investigate the powering mechanism of SN\,2020bqj by fitting the lightcurve with (a combination of) semi-analytical lightcurve models. We employ photometric modeling codes that have these models incorporated to provide fits to the SN lightcurve. First, we explore different models with \textsc{TigerFit} \citep{TigerFit}, where the best–fit model to a bolometric lightcurve of a transient is determined via $\chi^2$ minimization. Then, we use the Monte-Carlo code \textsc{MOSFiT} \citep{mosfit} to fit the full lightcurve across all filters with the model(s) that recovered the bolometric lightcurve best in the fits with \textsc{TigerFit}. As a Monte Carlo code, \textsc{MOSFiT} has the advantage of providing more robust statistical uncertainties, and also explicitly treats and fits for all variables (including, e.g., explosion epoch). Unlike \textsc{TigerFit} it also takes into account color information by fitting each band individually. However, it is significantly more computationally expensive, particularly for a data set the size of SN\,2020bqj.

\subsection{TigerFit}
\textsc{TigerFit} requires as input the phase in rest-frame since the explosion epoch, the bolometric luminosity (which we sample here at a daily cadence) and the associated uncertainty. For each model we fit for phases with discrete offsets $t_0$ from our adopted explosion epoch (MJD = 58880) of $t_0$ = [0, -1, -2, -3, -4, -5] days, and consider here the models with the lowest $\chi^2$ that still adhere to the upper limit on the luminosity from the last non-detection. We consider a radioactive decay model assuming diffusion into homologously expanding SN ejecta based on \citet{arnett1980,arnett1982}, and the CSM interaction model described in \citet{chatzopoulos2012,chatzopoulos2013}, which includes a large number of free parameters such as opacity $\kappa$ (assumed constant), ejecta mass $M_{\mathrm{ej}}$, CSM mass $M_{\mathrm{CSM}}$ and CSM density $\rho_{\mathrm{CSM}}$.  The ejecta density profile in the CSM model from \citet{chatzopoulos2012} is described separately for the inner and outer ejecta, as $\rho_{\mathrm{ej}} \propto r^{-\delta}$ and $\rho_{\mathrm{ej}} \propto r^{-n}$, respectively. We adopt $\delta = 1$ and $n = 11$, noting that the results are not very sensitive to these parameters. The density profile of the CSM is described as $\rho \propto r^{-s}$, where $s = 0$ corresponds to a CSM shell of constant density (the `shell' model), and $s = 2$ corresponds to a profile that better describes a wind (the `wind' model). We fit for both $s = 0$ and $s = 2$. 

Figure~\ref{fig:bolometric} shows the resulting model fits to the data, including the radioactive decay model fit, and hybrid model fits of radioactive decay plus either the CSM shell or wind model. Fits using only the CSM interaction model are not shown here, since they did not represent the data well using reasonable parameters\footnote{Manual tuning using only the CSM model required an ejecta mass of $\sim$200 M$_{\odot}$ to recover the lightcurve shape.}. The radioactive decay model to the full bolometric lightcurve does not capture the different stages of the rise, plateau and decline of the lightcurve (reduced $\chi^2$/d.o.f. = 5.9, when $t_0 = -2$), in particular not reproducing the sharp transitions between the three different stages. The Ni+CSM model fits recover the plateau much better, although depending on the CSM density profile the fits show different behaviours at early phases. Like the radioactive decay fits, the Ni+CSM fit with $s = 2$ rises monotonically, with the slope of the rise declining with time. The fit plotted in Fig.~\ref{fig:bolometric} assumes $t_0 = -3$ days (reduced $\chi^2$/d.o.f. = 0.8). In contrast with the smooth evolution of the other models, the Ni+CSM shell model shows sharp transitions between the rise, a (concave) plateau feature, and the decline. The upper limit before discovery constrains $t_0$ to $-4$ days, resulting in a reduced $\chi^2$/d.o.f. = 1.8 for the plotted fit. Both Ni+CSM model fits show a rise more rapid than the radioactive decay model, but neither are quite fast enough to fit the observed rise of SN\,2020bqj. 

While formally a slightly poorer fit, qualitatively the shell model with $s = 0$ recovers the shape of the bolometric lightcurve of SN\,2020bqj remarkably well, with a fast rise, a plateau connecting two peaks, followed by a linear decline. The sharp transitions in the shell model originate from the behaviour of the forward and reverse shock components that make up the Ni+CSM shell model. This is well demonstrated by \citet[][their figure 1]{wheeler2017}. The forward shock dictates the first half of the lightcurve, running through the CSM shell up until the transition between the rise and the (concave) plateau, when it reaches the outer edge of the CSM shell. The reverse shock dominates the later part of the lightcurve, with the transition between plateau and decline being the result of the shock reaching the interior of the ejecta. The following decline is a result of the diffusion from the shock-heated CSM matter. It is worth noting that the distinctive lightcurve features of SN\,2020bqj that are so well recovered by this Ni+CSM shell model were also observed in SN\,2011hw (Fig.~\ref{fig:template_comparison}), although its rise was not well constrained. The resulting parameters from fitting the bolometric lightcurve with the hybrid shell and wind CSM models in \textsc{TigerFit} are listed in Table~\ref{tab:tigerfit} for completeness, but should be considered secondary to the results obtained with \textsc{MOSFiT} in the following section.

From the \textsc{TigerFit} results we conclude the radioactive decay model does not reproduce the bolometric lightcurve well, and neither do models solely powered by CSM interaction. These models are thus rejected. Both hybrid Ni+CSM model fits (shell and wind) do recover the lightcurve well, although at early phases they show different behaviours. Next, we employ \textsc{MOSFiT} to fit the full multi-color lightcurve of SN\,2020bqj, instead of the pseudo-bolometric lightcurve, with both Ni+CSM models. By fitting the individual bands directly with \textsc{MOSFiT}, we are less sensitive to the assumptions that went into constructing the pseudo-bolometric lightcurve, such as the lack of evolution of the SED of the SN during the first 35 days. Here it must be noted that \textsc{MOSFiT} assumes a blackbody SED, which may be reasonable for SN\,2020bqj at early phases but less so after $\sim$15 days when the broader features, such as the typical SN Ibn pseudo-continuum, develop (Fig.~\ref{fig:spectra}).

\subsection{MOSFiT}
In the \textsc{MOSFiT} multi-band lightcurve fits of SN\,2020bqj with the Ni+CSM models, the values for $\delta$ and $n$ were adopted as above, and we allowed the progenitor radius and opacity to vary, resulting in a total of 12 free parameters in this model. We allowed all priors to vary over a wide range, regardless of the \textsc{TigerFit} results. As recommended for problems with a large number of free parameters, we use the dynamic nested sampler option in \textsc{MOSFiT} (implemented via the \textsc{dynesty} package; \citealt{Speagle2020a}). In either case, we initialize \textsc{MOSFiT} with 120 walkers, and allow it to run until it reaches the default stopping criterion in \textsc{dynesty} indicating that further sampling will no longer improve the evidence and posterior integrals (which took $\sim 28,000$ iterations in the case of $s=0$ and $\sim 31,000$ iterations for $s=2$).

\begin{figure*}
\centering
	\begin{tabular}{cc}
        \includegraphics[width=0.5\hsize]{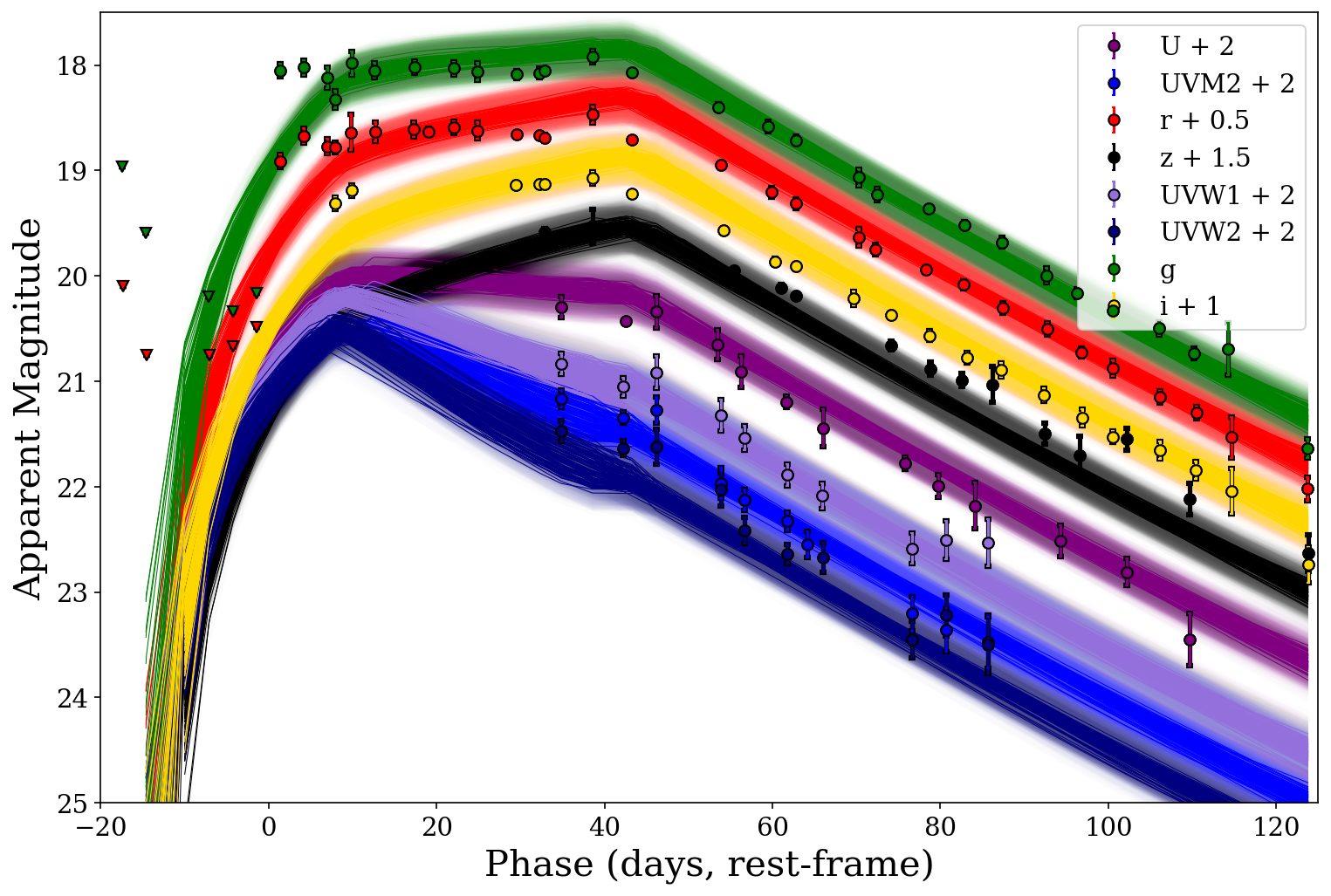}
        \includegraphics[width=0.5\hsize]{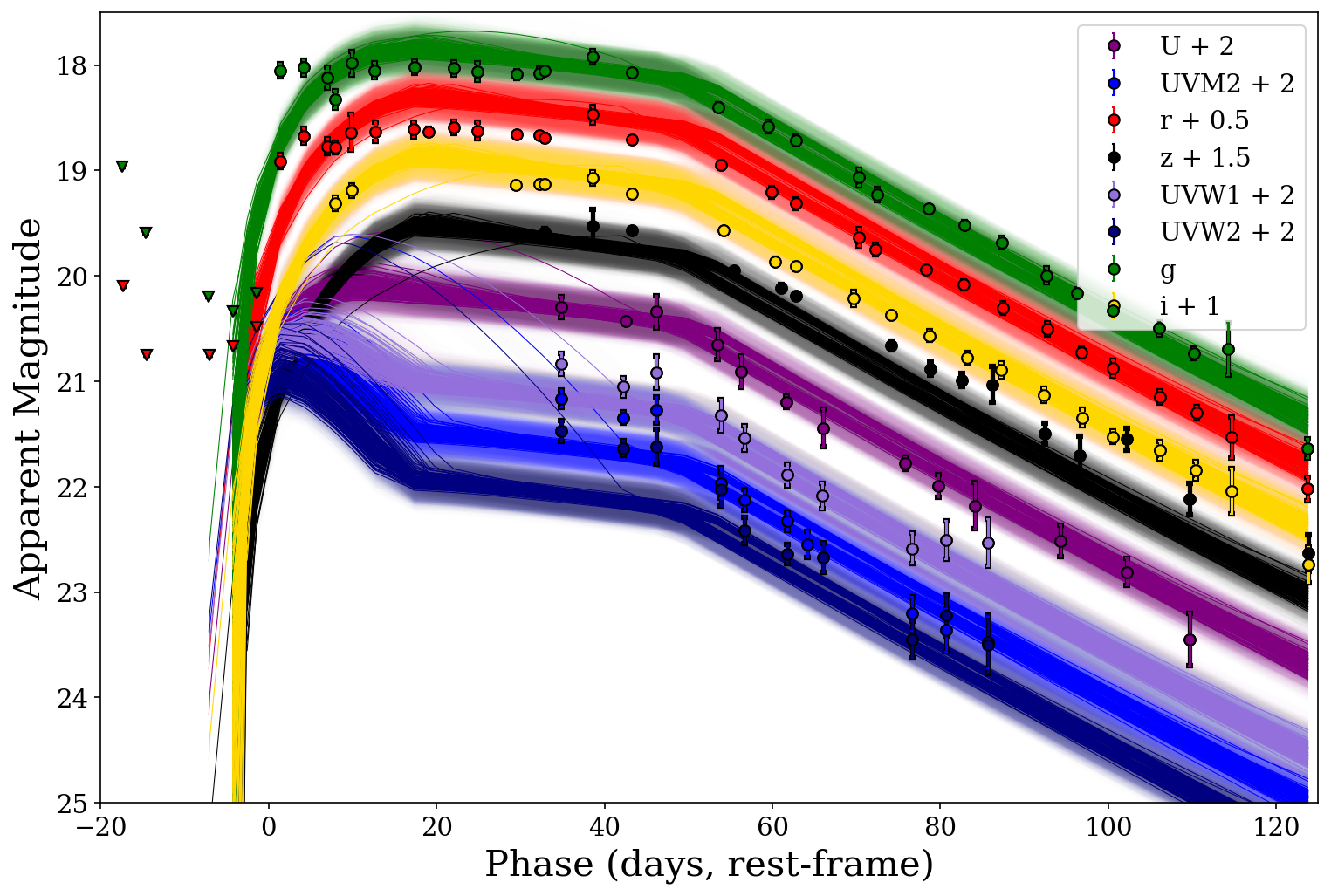}
    \end{tabular}
    \caption{Ni+CSM model lightcurves fitted to the photometry of SN\,2020bqj using the Monte-Carlo code \textsc{MOSFiT}, with on the left the shell model ($s = 0$) and on the right the wind model ($s = 2$). A selection of lightcurves based on a random draw from the posteriors are plotted, to demonstrate the range of the model fit in each filter. The models are able to reproduce the flat plateau in the optical bands as well as the steady decline phase of the UV/optical. The rapid rise to the peak plateau of SN\,2020bqj is less well reproduced, especially when taking into account the upper limits.}
\label{fig:mosfit}
\end{figure*}

The resulting lightcurve fits are shown in Fig.~\ref{fig:mosfit}. The prior ranges and best estimates of the fitted parameter values relevant to our discussion are listed in Table~\ref{tab:mosfit}. All listed parameters were well constrained by the data, with the uncertainties corresponding to the 68\% confidence interval in the posterior distributions. The four unlisted parameters include a noise parameter $\sigma$, which for both models was well constrained and of similar value ($\sigma\sim0.2$), and two `nuisance' parameters that are included to allow for a better determination of the more fundamental parameters such as ejecta mass and kinetic energy \citep{nicholl2017}. These nuisance parameters, the gamma-ray opacity of the SN ejecta $\kappa_{\gamma}$ and the H column density of the host $n_{\rm H, host}$ are not well constrained by the data. The last unlisted parameter is the final plateau temperature $T_{\rm f}$, to which the important physical parameters are not very sensitive \citep{nicholl2017}. The corner plots associated with the two model fits are shown in the Appendix (Fig.~\ref{fig:mosfit_corner} and Fig.~\ref{fig:mosfit_corner_s2}).

Both Ni+CSM models recover the plateau and decline phases of SN\,2020bqj quite well. Neither of the model lightcurves recover the rapid rise of SN\,2020bqj, in particular when accounting for the pre-discovery $g$- and $r$-band upper limits. Although upper limits are included in the fit, it is clear the optimal solution favours larger offsets of the free parameter $t_0$ than with \textsc{TigerFit}, where $t_0$ was fixed. We note that the upper limits reported here are global estimates of the full CCD \citep{masci2019}, and may differ locally. As a very recent SN, it is not possible at this point to improve this estimate for the position of SN\,2020bqj due to data access, but it is unlikely to change much given how faint the host of SN\,2020bqj is. We used the \texttt{dynesty} sampler \citep{Speagle2020a} in \textsc{MOSFiT} to sample the posterior probability function. The evaluation score, indicating how well a model fit the data, were similar for both models (103 and 107 for shell and wind model, respectively).

\begin{table*}
\centering
\begin{tabular}{lllcc} \hline
Parameter                       &Unit                   &Prior range        & \multicolumn{2}{c}{Realization} \\
                                &                       &               &$s = 0$                    &$s = 2$ \\  \hline \noalign{\vskip 5pt}
Nickel fraction ($f_{\rm Ni}$)  &\%                     &0.1 $-$ 100      &$0.24^{+0.19}_{-0.10}$     &$0.20^{+0.14}_{-0.07}$\\ [5pt]
CSM opacity ($\kappa_{\rm CSM}$)&cm$^2$ g$^{-1}$        &0.1 $-$ 0.34    &$0.25^{+0.05}_{-0.07}$     &$0.31^{+0.02}_{-0.05}$\\ [5pt]
Kinetic energy ($E_{\mathrm{kin}}$)&$10^{51}$ erg       &0.01 $-$ 20      &$1.6^{+0.1}_{-0.1}$     &$12.6^{+0.7}_{-1.3}$\\ [5pt]
CSM mass ($M_{\rm CSM}$)        &M$_{\odot}$            &0.01 $-$ 20       &$1.6^{+0.5}_{-0.3}$        &$0.02^{+0.02}_{-0.01}$\\ [5pt]
Ejecta mass ($M_{\rm ej}$)      &M$_{\odot}$            &1 $-$ 40         &$15.1^{+2.7}_{-1.7}$       &$16.6^{+2.1}_{-1.3}$\\ [5pt]
CSM inner radius ($R_0$)        &AU                     &0.001 $-$ 10      &$0.3^{+2.6}_{-0.3}$     &$0.08^{+0.12}_{-0.03}$\\ [5pt]
CSM density (log$_{10}$ $\rho_{\rm CSM}$)&g cm$^{-3}$   &-13 $-$ -6       &$-12.0^{+0.2}_{-0.1}$      &$-6.7^{+0.5}_{-0.8}$\\ [5pt]
Explosion epoch ($t_0$)         &days                   &-500 $-$ 0       &$-14.7^{+1.0}_{-1.2}$      &$-4.9^{+0.5}_{-0.6}$ \\  [5pt] \hline \noalign{\vskip 5pt}
Nickel mass ($M_{\rm Ni}$ = $f_{\rm Ni}\times M_{\rm ej}$)&M$_{\odot}$&$-$             &$0.04^{+0.04}_{-0.02}$     &$0.03^{+0.03}_{-0.01}$\\ [5pt]
Ejecta velocity ($v_{\mathrm{ej}} = \sqrt{2\times E_{\mathrm{kin}} / M_{\mathrm{ej}}}$) &km s$^{-1}$            &$-$             &$3300^{+300}_{-400}$       &$8700^{+600}_{-900}$\\ [5pt]
\hline
\end{tabular}
\caption{\label{tab:mosfit} Realized best estimates of eight free parameters used in fitting the hybrid Ni+CSM model to the multi-band lightcurve of SN\,2020bqj with \textsc{MOSFiT}, and three derived parameters. We considered the Ni+CSM shell model ($s = 0$) with a constant CSM density profile, and the wind ($s = 2$) model, with the CSM density dropping off as $1/r^2$. The explosion epoch $t_0$ (in the observer frame) is relative to our adopted explosion epoch. The listed uncertainties correspond to the 68\% confidence intervals in the posterior distributions (see the Appendix for the associated corner plots) and do not account for systematic uncertainties in the description of the data by the (simplified) models.}
\end{table*}

The first thing to note from the derived parameters in Table~\ref{tab:mosfit} is that the adopted density profile (wind or shell) of the CSM impacts some of the parameters significantly, while the produced lightcurves (Fig.~\ref{fig:mosfit}) are not very different. This dependency on index $s$ was already demonstrated by \citet{chatzopoulos2013} and gives us the main caveat of the semi-analytical CSM model we use here: the actual CSM configuration around the SN is uncertain, so the results should not be taken literally but rather indicative, especially when showing large variations as a function of the adopted density profile.

The parameters that seem insensitive to the adopted density profile are mainly associated with the progenitor properties. Both model fits result in similar ejecta (15 M$_{\odot}$ and 17 M$_{\odot}$) and nickel masses (0.04 M$_{\odot}$ and 0.03 M$_{\odot}$). The ejecta velocity\footnote{calculated as $v_{\rm ej} = \sqrt{2\times E_{\mathrm{kin}} / M_{\rm ej}}$.} is larger in the wind model (8700 km s$^{-1}$) as compared to the shell model (3300 km s$^{-1}$), as a result of the larger kinetic energy of the SN in the wind model. The radius $R_0$, which in this model corresponds to the inner radius of the CSM \citep{wheeler2017}, and the upper limit of the radius of the progenitor, is poorly constrained in the shell model, ranging up to 3 AU, or $\sim$ 600 R$_{\odot}$. Assuming a CSM velocity of $\sim$200 km s$^{-1}$, this implies the shell was ejected very recently before explosion (< a month). Such an outburst has not been observed at the position of SN\,2020bqj by ZTF, with coverage dating back two years \citep{strotjohann2020}. However, at the distance of SN\,2020bqj an outburst of similar brightness as that observed for the prototypical Type Ibn SN\,2006jc ($M = -14.0$ mag, \citealt{nakano2006}) would correspond to $m = 23$ mag, below the detection limit of nominal ZTF operations. The radius $R_0$ in the wind model is better constrained, and of order $\sim$0.1 AU, or 22 R$_{\odot}$.

The CSM masses show two orders of magnitude variation, and as such are much more uncertain. The large difference in CSM density $\rho_{\rm CSM}$ is expected, since it corresponds to the CSM density at the inner edge of the CSM at radius $R_0$. In the shell model the density remains constant, while in the wind model it drops off as $1/r^2$. The derived CSM properties are a good demonstration of the uncertainty in the CSM configuration. A constant CSM density as in the shell model of $10^{-12}$ g cm$^{-3}$ implies an electron density of $\sim10^{10}$ cm$^{-3}$, which is larger than observed in the line forming regions of Type IIn SNe ($\sim10^{5} - 10^{9}$ cm$^{-3}$, e.g., \citealt{fransson2014}). However, if the CSM originated from a wind, the mass-loss rate computed as $\dot{M} = 4\pi\ v_{\rm wind}\ R_{\rm CSM}^2\ \rho_{\rm CSM}$ results in an extremely high mass-loss rate of $\dot{M} \sim 1$ M$_{\odot}$ yr$^{-1}$, where we assume $v_{\textsc{wind}}$ = 230 km s$^{-1}$ inferred from the absorption velocity (Sect.~\ref{sec:spectra_analysis}). This mass-loss rate is much higher than typical mass-loss rates from steady-state winds of both LBV and WR stars ($\sim10^{-4}$ to $\sim10^{-5}$, \citealt{smith2017}) and rather suggests more violent episodic mass-loss or outbursts. The actual configuration of the CSM is almost certainly more complicated than the two density profiles considered here.

Finally, we note that the \textsc{MOSFiT} and \textsc{TigerFit} results are not fully consistent, which is likely a result of systematic uncertainties resulting from fundamental assumptions in the two different methods, such as the extrapolation to early times in \textsc{TigerFit} and the shape of the SED in \textsc{MOSFiT}. However, in either case a large ejecta mass and a high opacity are required to recover the lightcurve of SN\,2020bqj.

\section{Discussion} \label{sec:discussion}
SN\,2020bqj is a clear outlier in the SN Ibn subclass, based on the combination of its unusual lightcurve, spectral features and host galaxy properties. Previous studies of `normal' fast-evolving SNe Ibn have employed similar methods and models to model SN Ibn lightcurves as we have, which allows us to draw direct comparisons of our derived model parameters with their work. The ejecta mass of $15 - 17$ M$_{\odot}$ of SN\,2020bqj inferred from the Ni+CSM models is similar to those derived for Type Ibn SN\,2019uo \citep{gangopadhyay2020} and PS15dpn \citep{wang2019}. The $^{56}$Ni mass of $0.03 - 0.04$ M$_{\odot}$ inferred from the modeling is comparable to the values for SN\,2019uo and PS15dpn, as well as for the rapidly-evolving SE SN with narrow He features LSQ13ddu \citep{clark2020}, and is consistent with the amount of $^{56}$Ni synthesized in the explosions of normal Type Ib/c SE SNe \citep[e.g.,][]{prentice2016}. Since the ejecta and $^{56}$Ni masses are consistent with SNe Ibn with fast monotonically evolving lightcurves, it seems unlikely the explosion properties are the origin of the unusual lightcurve of SN\,2020bqj. 

Where the best fit model parameters of SN\,2020bqj differ from those of the aforementioned SNe Ibn is in the properties of the CSM. The inner radius of the CSM, $R_0$, of $\sim$0.1$ - $0.3 AU ($\sim$22 R$_{\odot}$) is orders of magnitudes smaller than the radii derived for SN\,2019uo, PS15dpn and LSQ13ddu. Also the inferred CSM optical opacity of $\kappa\sim0.3$ cm$^{2}$ g$^{-1}$ is higher for SN\,2020bqj than in the model fits of these fast-evolving SNe Ibn (0.05 - 0.1 cm$^{2}$ g$^{-1}$). Interestingly, the lightcurve of the long-lived SN Ibn OGLE-2014-SN-131 (Fig.~\ref{fig:template_comparison}) was well represented by a CSM shell model with higher opacity of $\kappa = 0.2$ cm$^{2}$ g$^{-1}$ and a small radius of $R$ = 1 R$_{\odot}$ \citep{karamehmetoglu2017}. Such an elevated opacity could be explained if the CSM not only contains He but also H (typical opacity $\kappa\sim 0.34$ cm$^{2}$ g$^{-1}$). The presence of H in the CSM is supported by the (albeit faint) detection of Balmer lines in the spectra of OGLE-2014-SN-131, and they are also observed in SN\,2020bqj (Fig.~\ref{fig:line_evolution}). In fact, an elevated H opacity was hypothesized by \citet{smith2012} to explain the properties of the peculiar SN Ibn SN\,2011hw, which seems to be an analog to SN\,2020bqj both in terms of its lightcurve shape (Fig.~\ref{fig:template_comparison}) and the evolution of its spectra (Fig.~\ref{fig:line_identification}). SN\,2011hw was a well-studied event discussed by both \citet{smith2012} and \citet{pastorello2015IV}, and we consider here their interpretation of the properties of SN\,2011hw in the context of SN\,2020bqj.

\citet{smith2012} compared the photometric and spectral properties of SN\,2011hw to those of the prototypical Type Ibn SN\,2006jc. The presence of more prominent Balmer lines and a 6000 K blackbody component (reminiscent of Type IIn SNe) in late time spectra of SN\,2011hw, when compared to SN\,2006jc, led \citet{smith2012} to hypothesize that the main difference between SN\,2011hw and SN\,2006jc could be due to a slightly enhanced H-abundance in the CSM and the progenitor envelope of SN\,2011hw. The stronger continuum H opacity would explain the higher luminosity and slower decline rate at late times of SN\,2011hw as compared to SN\,2006jc. The later appearance of Ca\,{\sc ii} (day 47 for SN\,2011hw versus day 13 for SN\,2006jc, \citealt{foley2007}) would then be a result of the higher optical depths at large radii due to the elevated H abundance, hiding Ca\,{\sc ii} emission lines arising from the SN ejecta crossing the reverse shock. In comparison, the shape of the continuum and the H$\alpha$ line strength of SN\,2020bqj matches closely that for SN\,2011hw (Fig.~\ref{fig:line_identification}). The appearance of Ca\,{\sc ii} at +71 days is also delayed for SN\,2020bqj (Fig.~\ref{fig:spectra}), even with respect to SN\,2011hw. SN\,2020bqj also shows the same distinctive lightcurve properties as SN\,2011hw, and arguably more prominent. SN\,2020bqj is brighter than SN\,2011hw by $\sim$1 magnitude (Fig.~\ref{fig:template_comparison}), and its linear decline at 0.04 mag day$^{-1}$ is slower than for SN\,2011hw (0.055 mag day$^{-1}$, \citealt{pastorello2015IV}).

The re-brightening of SN\,2011hw at the end of the plateau was interpreted as additional luminosity input from the shock running into a denser portion of its CSM shell \citep{smith2012,pastorello2015IV}. In the case of SN\,2020bqj a similar secondary peak is well recovered by the Ni+CSM model (most prominently shown by the \textsc{TigerFit} shell model fit, Fig.~\ref{fig:bolometric}) as the transition from the (concave) plateau phase to the decline phase due to the reverse shock reaching the interior of the ejecta, after which the lightcurve declines, dominated by the diffusion from the reverse shock heated matter \citep{wheeler2017}. The lightcurve of SN\,2011hw, where observed, is similar to that of SN\,2020bqj in the plateau and decline phase. Therefore, we suggest that the light curve of SN\,2011hw is dominated by shock heating in a scenario similar to that described for SN\,2020bqj, rather than variations in the CSM density. Arguably this scenario could also be applied to the lightcurve of iPTF13beo, which showed a fast rise, double peak, and a subsequent decline (Fig.~\ref{fig:template_comparison}). The faster evolution of iPTF13beo compared to SN\,2011hw and SN\,2020bqj would then be a result of less H in the CSM, which is consistent with the lack of Balmer lines in the spectra of iPTF13beo \citep{gorbikov2014}.

The elevated H abundance in the CSM of SN\,2020bqj should reflect the composition of the envelope of the progenitor, since the CSM is assumed to have originated from the progenitor through mass-loss. The spectra of SN\,2011hw showed narrow blue-shifted (by $-80$ to $-250$ km s$^{-1}$) P-Cygni absorption features superimposed on strong emission lines. These absorption velocities correspond to the pre-shocked velocity of the CSM, and are unusually slow in SN\,2011hw when compared to typical pre-shock CSM velocities of SNe Ibn ($\sim$1000 km s$^{-1}$, \citealt{pastorello2016}). \citet{smith2012} argued that the slower absorption velocities reflect a larger radius and thus lower escape velocity of its progenitor star than expected from classic, H-depleted WR stars (500 - 3200 km s$^{-1}$, \citealt{crowther2007}), implying a progenitor with a higher H content such as an LBV (< 500 km s$^{-1}$), which would be more consistent with a elevated H abundance in the CSM. Notably, Type Ibn SN\,2005la \citep{pastorello2008II}, which had a peculiar fluctuating lightcurve (Fig.~\ref{fig:template_comparison}), also showed slow unshocked CSM velocities ($\sim500$ km s$^{-1}$), as well as prominent Balmer lines. In SN\,2020bqj we detected absorption features of $\lesssim$230 km s$^{-1}$ in the high SNR spectrum obtained at +47 days (Sect.~\ref{sec:spectra_analysis}), which are consistent with mass-loss from an LBV star rather than a compact WR star. The inner radius of the CSM of SN\,2020bqj inferred from the wind model ($s = 2$) of $R_0 = 17^{+26}_{-7}$ R$_{\odot}$ can be considered an upper limit to the progenitor radius. Such a small radius would favour a more compact progenitor such as a WR star \citep{petrovic2006}, but given the large uncertainties, it does not exclude an LBV star \citep[e.g.,][]{Sholukhova2015}.

Thus, the observed higher luminosity, slow wind speeds, slower decline rate, and delayed emergence of Ca\,{\sc ii} in SN\,2020bqj would suggest a CSM with an elevated H-opacity compared to fast evolving SNe Ibn, similar to SN\,2011hw, but potentially with an higher H-abundance. An elevated opacity is consistent with the model fits of the lightcurve of SN\,2020bqj, where an opacity $\kappa_{\rm CSM}$ was derived of 0.25 -- 0.31 g cm$^{-3}$, depending on the model. \citet{smith2012} and \citep{pastorello2015IV} suggested as progenitor options for SN\,2011hw a massive star transitioning from the LBV to WR stage: either a early WN star with H or a member of the Ofpe/WN9 class of stars, which are very massive stars with ZAMS masses of $17-100$ M$_{\odot}$ \citep{stlouis1997}. Such a massive progenitor, at a stage in its evolution between LBV and WR star where it has not yet lost most of its envelope through mass-loss, is consistent with the large ejecta mass of $\sim15 - 17$ M$_{\odot}$ inferred for SN\,2020bqj. In contrast, normal SNe Ib/c (the underlying SNe assumed for most SNe Ibn) are linked to evolved WR progenitors and have ejecta masses of order a few M$_{\odot}$ \citep{drout2011,lyman2016,prentice2016,Barbarino2020}. A scenario of a progenitor with LBV-like properties also suits the lightcurve and host galaxy of SN\,2020bqj, since LBV stars have been linked to Type IIn SNe \citep[e.g.,][]{galyam2009,smith2010}. Type IIn SNe display a variety of (long-lasting) lightcurves including SNe with plateaus similar to SN\,2020bqj, such as PTF11rfr (Fig.~\ref{fig:template_comparison}; \citealt{nyholm2020}). Additionally, the host galaxy of SN\,2020bqj falls well within the distribution in host galaxy masses of Type IIn SNe \citep{schulze2020}, whereas SN\,2020bqj is an outlier in host galaxy mass compared to SNe Ibn (Sect.~\ref{sec:host_analysis}). Here it must be noted that due to the rarity of SNe Ibn the iPTF sample is limited in size (9 objects), and the uncertainty on the host galaxy mass of SN\,2020bqj is large, which makes it unclear if the host of SN\,2020bqj could still be part of a continuous mass distribution.

\citet{pastorello2015IV} put forward the suggestion that SNe Ibn are objects in a sequence between events related to LBV stars, such as (some) SNe IIn, and evolved WR stars, such as (some) SNe Ib/c. Based on the properties of the progenitor and the related CSM inferred from the spectra and lightcurve, SN\,2020bqj in this scheme would occupy a different space than the typical fast-evolving H-deficient SNe Ibn. Like SN\,2011hw, SN\,2020bqj would be considered a transitional SN Ibn/IIn resulting from a WR/LBV-like progenitor, producing a relatively opaque H-rich CSM (compared to the CSM in normal SNe Ibn), whereas regular rapidly evolving SNe Ibn would result from more evolved H-deficient WR-like progenitors. While this scheme appears to offer a plausible explanation for why we have observed two SNe Ibn that are very similar, but also distinctively different from the main population of SNe Ibn, it is hampered by the limited sample size of these transitional SN Ibn/IIn (where we note that this classification extends beyond just spectral properties). Additionally, this interpretation does not explain why some rapid evolving SNe Ibn still show prominent H \citep[e.g. SN\,2018bcc;][]{karamehmetoglu2019} in their spectra, but a slow evolving SN Ibn like OGLE-2014-SN-13 only shows very little H.

\section{Summary and conclusions} \label{sec:summary}
While the spectral features, early lightcurve evolution and peak magnitude firmly establish SN\,2020bqj as a SN Ibn, in contrast with the typical members of this subclass SN\,2020bqj remained at peak brightness for $\sim$40 days, declined only slowly afterwards, and showed little spectral evolution throughout. Based on the observational properties of SN\,2020bqj, the modeling of the multi-band lightcurve with \textsc{MOSFiT}, and a comparison with the strikingly similar Type Ibn SN\,2011hw, we propose that SN\,2020bqj is the result of a transitional LBV/WR progenitor exploding in a dense He-rich CSM with an elevated H-opacity. This conclusion is based on the following:

\begin{itemize}
\setlength\itemsep{1em}
	\item The distinctive phases of the lightcurve are recovered well by a model where the luminosity input is dominated by reverse and forward shock heating from the interaction of the SN ejecta and CSM, and a minor contribution by $^{56}$Ni decay.
	
	\item Based on the spectra the CSM around SN\,2020bqj is He-rich, but also contains H. The presence of H increases the optical opacity of the CSM. An elevated CSM opacity is also suggested by the model fits, and is likely the leading cause for the long-lived and luminous lightcurve of SN\,2020bqj.
	
	\item The ejecta mass of SN\,2020bqj derived from the modeling of $\sim15 - 17$ M$_{\odot}$ is consistent with a very massive progenitor. The low observed pre-shock absorption velocity implies a progenitor with a larger stellar radius. As for SN\,2011hw, this implies a progenitor such as a post-LBV star with H in the envelope, rather than a compact H-depleted late-type WR star.
\end{itemize}

SN\,2020bqj and SN\,2011hw seem to represent a transitional subclass between Type Ibn and Type IIn SNe. They show not only prominent He lines, but also residual H in their spectra, have lightcurves that are much longer-lived than typical SNe Ibn, and show features that are consistent with progenitors with both LBV and WR-like properties. The discovery of SN\,2020bqj, which is so similar to SN\,2011hw, supports the notion \citep{smith2012,pastorello2015IV} that such SNe represent a short-lived but distinct phase in the stellar evolution of massive stars between the LBV and WR phases, and provides evidence of a continuity between Type Ibn and IIn SNe. SN\,2020bqj and SN\,2011hw are outliers in the already rare subclass of SNe Ibn, and there should be an observational bias in favour of detecting longer-lived bright SNe such as SN\,2020bqj. This rarity could be a reflection of the short duration of the stellar evolution phase of their progenitors, or the small probability for such stars to undergo core collapse. It would require a larger sample of SNe Ibn to further study this potential distinct subclass of shock-dominated Type Ibn/IIn SNe. Fortunately current widefield synoptic surveys such as ZTF, Pan-STARRS \citep{chambers2016}, the Asteroid Terrestrial-impact Last Alert System \citep[ATLAS;][]{tonry2018,smith2020}, and the upcoming deluge of transient detections expected from the Legacy Survey of Space and Time \citep[LSST;][]{ivezic2019} should dramatically increase the SN Ibn sample, and all rare sub-types with it.

\begin{acknowledgements}
We thank the anonymous referee for their comprehensive feedback and suggestions, which has improved this work. ECK acknowledges support from the G.R.E.A.T research environment funded by {\em Vetenskapsr\aa det}, the Swedish Research Council, under project number 2016-06012, and support from The Wenner-Gren Foundations. RL is supported by a Marie Sk\l{}odowska-Curie Individual Fellowship within the Horizon 2020 European Union (EU) Framework Programme for Research and Innovation (H2020-MSCA-IF-2017-794467). ESP's research was funded in part by the Gordon and Betty Moore Foundation through Grant GBMF5076. Y-LK has received funding from the European Research Council (ERC) under the European Unions Horizon 2020 research and innovation program (grant agreement No. 759194 USNAC). MLG acknowledges support from the DiRAC Institute in the Department of Astronomy at the University of Washington. AH is grateful for the support by grants from the Israel Science Foundation, the US-Israel Binational Science Foundation, and the I-CORE Program of the Planning and Budgeting Committee and the Israel Science Foundation.

Based on observations obtained with the Samuel Oschin 48-inch Telescope and the 60-inch Telescope at the Palomar Observatory as part of the Zwicky Transient Facility project, a  scientific  collaboration  among  the  California  Institute  of Technology,  the  Oskar  Klein  Centre,  the  Weizmann  Institute of Science, the University of Maryland, the University of Washington, Deutsches Elektronen-Synchrotron, the University of Wisconsin-Milwaukee, and the TANGO Program of the University System of Taiwan.  Further support is provided by the U.S. National Science Foundation under Grant No. AST-1440341.

This work is based in part on observations made with the Nordic Optical Telescope, operated by the Nordic Optical Telescope Scientific Association at the Observatorio del Roque de los Muchachos, La Palma, Spain, of the Instituto de Astrofisica de Canarias.

This research uses services or data provided by the Astro Data Lab at NSF's National Optical-Infrared Astronomy Research Laboratory. NSF's OIR Lab is operated by the Association of Universities for Research in Astronomy (AURA), Inc. under a cooperative agreement with the National Science Foundation. 
 
The Photometric Redshifts for the Legacy Surveys (PRLS) catalog used in this paper was produced thanks to funding from the U.S. Department of Energy Office of Science, Office of High Energy Physics via grant DE-SC0007914.

The data presented here were obtained in part with ALFOSC, which is provided by the Instituto de Astrofisica de Andalucia (IAA-CSIC) under a joint agreement with the University of Copenhagen and NOTSA

The Liverpool Telescope is operated on the island of La Palma by Liverpool John Moores University in the Spanish Observatorio del Roque de los Muchachos of the Instituto de Astrofisica de Canarias with financial support from the UK Science and Technology Facilities Council.

SED Machine is based upon work supported by the National Science Foundation under Grant No. 1106171.

This work was supported by the GROWTH project funded by the National Science Foundation under Grant No 1545949.

The ZTF forced-photometry service was funded under the Heising-Simons Foundation grant \#12540303 (PI: Graham). 

We would like to thank participating observers on the UW APO ZTF follow-up team, including: Brigitta Spi\H{o}cz, Keaton Bell, and James Davenport. The DiRAC Institute is supported through generous gifts from the Charles and Lisa Simonyi Fund for Arts and Sciences, and the Washington Research Foundation.

We thank the staff of the Mullard Radio Astronomy Observatory, University of Cambridge, for their support in the maintenance, and operation of AMI. We acknowledge support from the European Research Council under grant ERC-2012-StG-307215 LODESTONE.

We thank David Kaplan and Eran Ofek for careful reading and suggestions.

\end{acknowledgements}
\bibliography{main}

\begin{appendix}
\section{Photometry and spectroscopy}

\begin{table*}
\caption{\label{tab:photom}Optical and ultraviolet photometry of SN\,2020bqj, in observed magnitudes. Phase is relative to estimated explosion epoch, in rest-frame.}
\centering
\begin{tabular}{ccccccc} \hline
        MJD &   Phase & Filter &     magnitude & error & Telescope & Instrument \\ \hline
  58853.54 &  -24.9 &      g &     >20.3 &                 &       P48 &        ZTF \\
 58856.55 &  -22.0 &      g &     >20.4 &                 &       P48 &        ZTF \\
 58856.57 &  -22.0 &      r &     >20.2 &                 &       P48 &        ZTF \\
 58861.49 &  -17.4 &      g &     >19.0 &                 &       P48 &        ZTF \\
 58861.55 &  -17.3 &      r &     >19.6 &                 &       P48 &        ZTF \\
 58864.46 &  -14.6 &      g &     >19.6 &                 &       P48 &        ZTF \\
 58864.52 &  -14.6 &      r &     >20.2 &                 &       P48 &        ZTF \\
 58872.45 &   -7.1 &      g &     >20.2 &                 &       P48 &        ZTF \\
 58872.56 &   -7.0 &      r &     >20.2 &                 &       P48 &        ZTF \\
 58875.48 &   -4.3 &      g &     >20.3 &                 &       P48 &        ZTF \\
 58875.54 &   -4.2 &      r &     >20.2 &                 &       P48 &        ZTF \\
 58878.54 &   -1.4 &      g &     >20.2 &                 &       P48 &        ZTF \\
 58878.56 &   -1.4 &      r &     >20.0 &                 &       P48 &        ZTF \\
 58881.52 &    1.4 &      g &     18.05 &            0.06 &       P48 &        ZTF \\
 58881.55 &    1.4 &      r &     18.41 &            0.06 &       P48 &        ZTF \\
 58884.47 &    4.2 &      g &     18.02 &            0.07 &       P48 &        ZTF \\
 58884.54 &    4.2 &      r &     18.17 &            0.07 &       P48 &        ZTF \\
 58887.46 &    7.0 &      g &     18.12 &            0.10 &       P48 &        ZTF \\
 58887.53 &    7.0 &      r &     18.27 &            0.07 &       P48 &        ZTF \\
 58888.47 &    7.9 &      r &     18.28 &            0.05 &       P60 &       SEDM \\
 58888.47 &    7.9 &      g &     18.32 &            0.08 &       P60 &       SEDM \\
 58888.47 &    7.9 &      i &     18.31 &            0.06 &       P60 &       SEDM \\
 58890.50 &    9.8 &      r &     18.14 &            0.17 &       P48 &        ZTF \\
 58890.55 &    9.9 &      g &     17.98 &            0.11 &       P48 &        ZTF \\
 58890.56 &    9.9 &      i &     18.19 &            0.06 &       P60 &       SEDM \\
 58893.50 &   12.6 &      g &     18.05 &            0.07 &       P48 &        ZTF \\
 58893.53 &   12.7 &      r &     18.13 &            0.09 &       P48 &        ZTF \\
 58898.50 &   17.3 &      r &     18.11 &            0.07 &       P48 &        ZTF \\
 58898.56 &   17.4 &      g &     18.02 &            0.06 &       P48 &        ZTF \\
 58900.35 &   19.1 &      r &     18.13 &            0.03 &       P60 &       SEDM \\
 58903.52 &   22.0 &      r &     18.09 &            0.06 &       P48 &        ZTF \\
 58903.53 &   22.0 &      g &     18.03 &            0.06 &       P48 &        ZTF \\
 58906.52 &   24.8 &      r &     18.12 &            0.08 &       P48 &        ZTF \\
 58906.53 &   24.8 &      g &     18.06 &            0.08 &       P48 &        ZTF \\
 58911.46 &   29.5 &      r &     18.16 &            0.01 &       P60 &       SEDM \\
 58911.46 &   29.5 &      g &     18.08 &            0.05 &       P60 &       SEDM \\
 58911.46 &   29.5 &      i &     18.14 &            0.03 &       P60 &       SEDM \\
 58911.52 &   29.5 &      r &     18.16 &            0.08 &       P48 &        ZTF \\
 58911.53 &   29.5 &      g &     18.13 &            0.08 &       P48 &        ZTF \\
 58911.55 &   29.6 &      g &     18.07 &            0.08 &       P48 &        ZTF \\
 58914.34 &   32.2 &      r &      18.20 &            0.04 &       P60 &       SEDM \\
 58914.37 &   32.2 &      r &     18.16 &            0.02 &       P60 &       SEDM \\
 58914.37 &   32.2 &      g &     18.08 &            0.05 &       P60 &       SEDM \\
 58914.37 &   32.2 &      i &     18.13 &            0.03 &       P60 &       SEDM \\
 58914.48 &   32.3 &      g &     18.06 &            0.08 &       P48 &        ZTF \\
 58914.52 &   32.3 &      r &     18.15 &            0.07 &       P48 &        ZTF \\
 58915.08 &   32.9 &      z &     18.08 &            0.03 &        LT &        IO:O \\
 58915.08 &   32.9 &      i &     18.13 &            0.02 &        LT &        IO:O \\
 58915.08 &   32.9 &      g &     18.05 &            0.02 &        LT &        IO:O \\
 58915.08 &   32.9 &      r &     18.19 &            0.03 &        LT &        IO:O \\
 58917.22 &   34.9 &   UVW1 &     18.83 &            0.10 &     Swift &       UVOT \\
 58917.22 &   34.9 &      U &      18.30 &            0.10 &     Swift &       UVOT \\
 58917.22 &   34.9 &      B &     17.95 &            0.11 &     Swift &       UVOT \\
 58917.22 &   34.9 &   UVW2 &     19.47 &            0.10 &     Swift &       UVOT \\
 58917.22 &   34.9 &      V &     18.01 &            0.22 &     Swift &       UVOT \\
 58917.23 &   34.9 &   UVM2 &     19.16 &            0.08 &     Swift &       UVOT \\
 58921.13 &   38.6 &      g &     17.92 &            0.06 &        LT &        IO:O \\
 58921.13 &   38.6 &      r &     17.97 &            0.08 &        LT &        IO:O \\
 58921.14 &   38.6 &      z &     18.03 &            0.16 &        LT &        IO:O \\ \hline
\multicolumn{7}{c}{ } \\
\multicolumn{7}{c}{(This table is available online in its entirety.)}
\end{tabular}
\end{table*}

\begin{table*}
\caption{\label{tab:spectral_log}Log of spectroscopic observations of SN\,2020bqj. Phase is relative to estimated explosion epoch, in rest-frame.}
\centering
\begin{tabular}{|c|c|r|c|c|c|c|c|c|c|} \hline
UT Date & MJD &Phase&Telescope & Instrument &  Range & Resolving power\\ \hline
 &  &(days)& &  &(\AA)&\\ \hline
2020-02-06.5 &  58885.5  &+5     &P60    &SEDM   & 3780$-$9220  &100  \\
2020-02-09.1 &  58888.1  &+8     &LT     &SPRAT  & 4020$-$7990  &350  \\
2020-02-11.4 &  58890.4  &+10     &P60    &SEDM   & 3780$-$9220 &$\sim$100  \\
2020-02-17.2 &  58896.2  &+15    &LT     &SPRAT  & 4020$-$7990  &350  \\
2020-02-17.5 &  58896.5  &+15    &APO    &DIS    & 5400$-$9000  &$\sim$700  \\
2020-02-21.4 &  58900.4  &+19    &P60    &SEDM   & 3780$-$9220  &$\sim$100  \\
2020-02-25.2 &  58904.2  &+23    &NOT    &ALFOSC & 3520$-$9640  &360  \\
2020-03-06.3 &  58914.3  &+32    &P60    &SEDM   & 3780$-$9220  &$\sim$100  \\
2020-03-10.1 &  58918.1  &+36    &NOT    &ALFOSC & 3800$-$9640  &360  \\
2020-03-22.5 &  58930.5  &+47    &Keck   &LRIS   & 3060$-$10310 &$\sim$900  \\
2020-03-28.5 &  58936.5  &+53    &P60    &SEDM   & 3780$-$9220  &$\sim$100  \\
2020-04-01.1 &  58940.1  &+56    &LT     &SPRAT  & 4020$-$7990  &350  \\
2020-04-04.5 &  58943.5  &+60    &P60    &SEDM   & 3780$-$9220  &$\sim$100  \\
2020-04-17.2 &  58956.2  &+71    &NOT    &ALFOSC & 3800$-$9640  &360  \\
2020-05-07.0 &  58976.0  &+90    &NOT    &ALFOSC & 3800$-$9640  &360  \\
2020-05-30.0 &  58999.0  &+112    &NOT    &ALFOSC & 3800$-$9640 &360  \\
\hline
\end{tabular}
\end{table*}

\section{TigerFit results}
\begin{table}
\centering
\begin{tabular}{llcc} \hline
Parameter                       &Unit   & \multicolumn{2}{c}{Realization} \\
                                &          &$s = 0$                    &$s = 2$ \\  \hline \noalign{\vskip 5pt} 
Nickel mass ($M_{\rm Ni}$)      &M$_{\odot}$        &0.30   &0.0001   \\ [5pt]
CSM opacity ($\kappa_{\rm CSM}$)&cm$^2$ g$^{-1}$    &0.24   &0.4   \\ [5pt]
Kinetic energy ($E_{\mathrm{kin}}$)&$10^{51}$ erg   &0.73   &0.54   \\ [5pt]
CSM mass ($M_{\rm CSM}$)        &M$_{\odot}$        &0.59   &1.4   \\ [5pt]
Ejecta mass ($M_{\rm ej}$)      &M$_{\odot}$        &13.3   &8.6   \\ [5pt]
CSM inner radius ($R_0$)        &AU                 &1.9   &0.007   \\ [5pt]
Explosion epoch ($t_0$)         &days               &-4     &-3   \\ [5pt]
$\chi^2$                        &                   &1.81   &0.83   \\ [5pt]
\hline
\end{tabular}
\caption{\label{tab:tigerfit} Resulting parameters from fitting the pseudo-bolometric lightcurve of SN\,2020bqj with \textsc{TigerFit}. The (underestimated) statistical errors associated with the $\chi^2$-minimization fit are not included, since they were not meaningful.}
\end{table}

\section{MOSFiT corner plots}

\begin{figure*}
\centering
    \includegraphics[width=0.95\hsize]{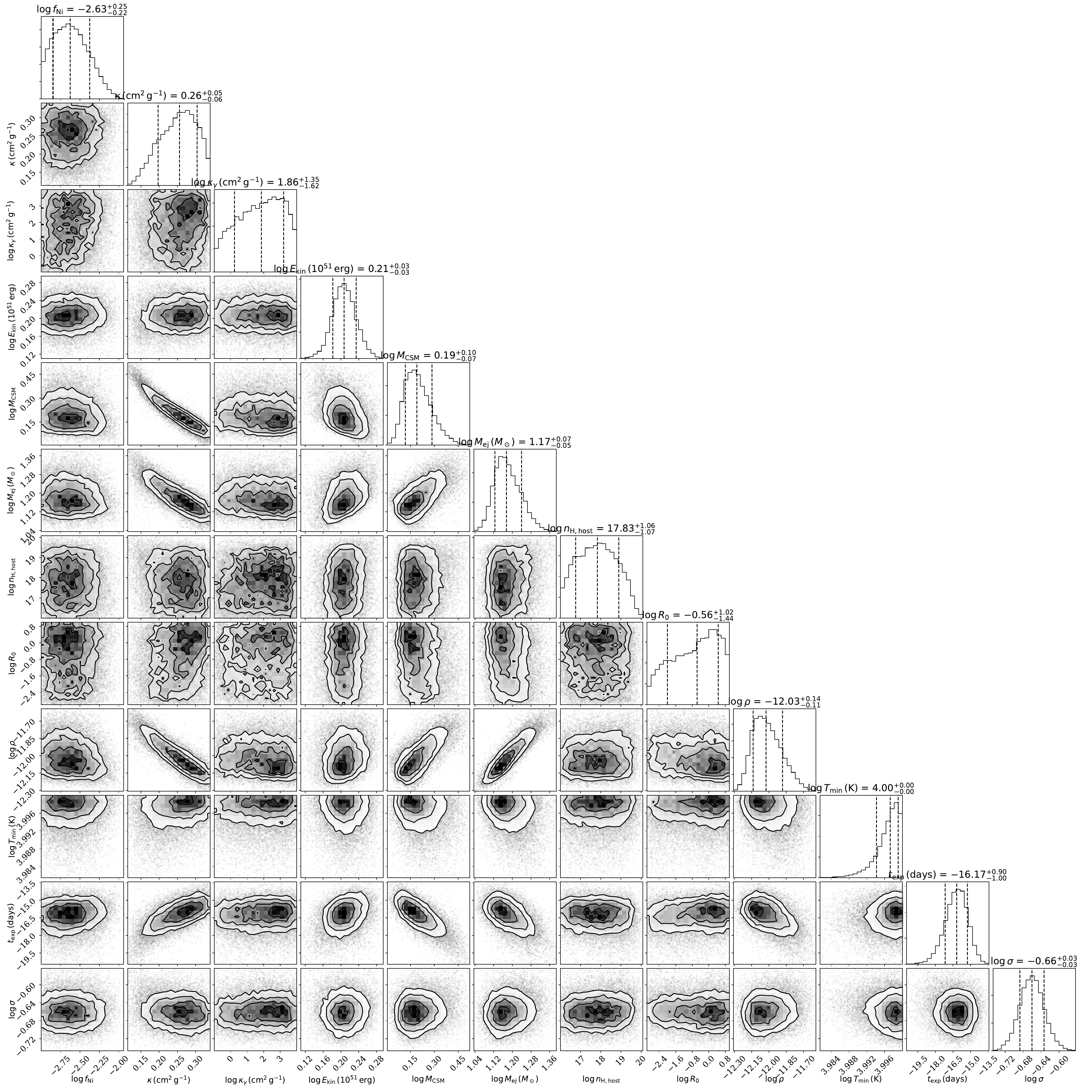}
    \caption{Corner plot of the parameter realization distributions of the Ni+CSM shell model fit (with $s = 0$) with \textsc{MOSFiT}. The fitted values are indicated by the center lines, with the 68\% confidence interval indicated by the lines on either side. Note that the parameters were fitted in log space.}
    \label{fig:mosfit_corner}
\end{figure*}

\begin{figure*}
\centering
    \includegraphics[width=0.95\hsize]{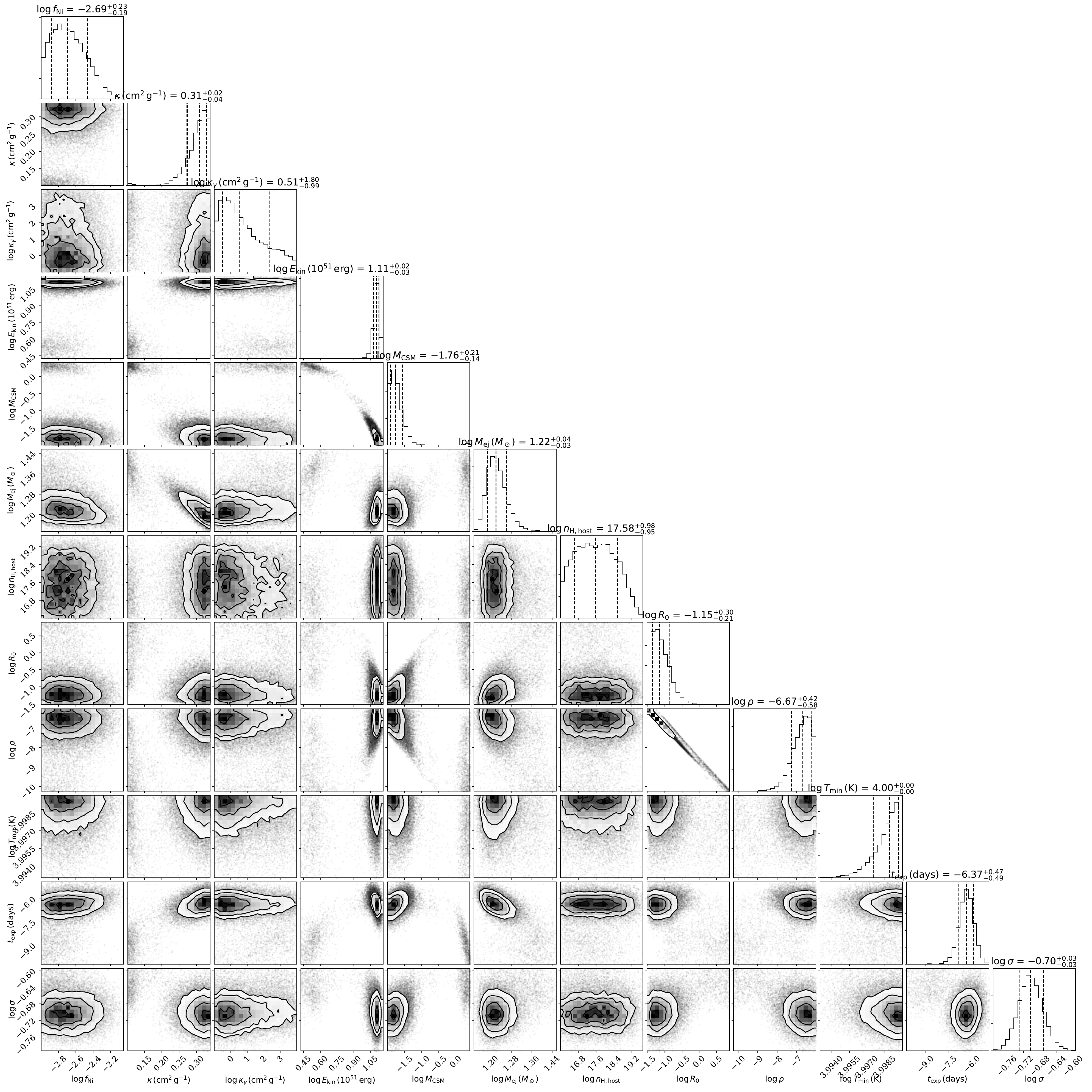}
    \caption{Corner plot of the parameter realization distributions of the Ni+CSM wind model fit (with $s = 2$) with \textsc{MOSFiT}. The fitted values are indicated by the center lines, with the 68\% confidence interval indicated by the lines on either side. Note that the parameters were fitted in log space.}
    \label{fig:mosfit_corner_s2}
\end{figure*}

\end{appendix}

\end{document}